\documentstyle[12pt,epsfig]{article}
\input epsf

\begin{document}

\title{Charmonium Production in Nucleus-Nucleus Collisions}

\author{Berndt M\"uller\\
Department of Physics, Duke University, Durham, NC 27708-0305}

\date{\today}

\maketitle

\begin{abstract}
This is a review of theoretical attempts to describe production of heavy
quark bound states in nucleus-nucleus collisions, in particular, the
relative suppression of $J/\psi$ and $\psi'$ production observed in
these reactions.  The review begins with a survey of experimental data
for proton-induced reactions and their theoretical interpretation.  The
evidence for additional suppression in nucleus-nucleus collisions is
discussed and various theoretical models of charmonium absorption by
comoving matter are presented and analyzed.  The review concludes with
suggestions for future research that would help clarify the
implications of $J/\psi$ suppression in Pb + Pb collisions observed by
the NA50 experiment.

\end{abstract}

\baselineskip=18pt

\section{Introduction}

Ever since their discovery two decades ago, bound states of
heavy quarks have served as one of the primary probes of quantum
chromodynamics.  The spectroscopy of charmonium and bottomonium states
provided crucial tests for the theory of forces among quarks,
culminating in recent years in the precise determination of the
strong coupling constant $\alpha_s$ by means of lattice simulations.

Because of their relatively small size, bound states of heavy quarks
interact only rather weakly with other hadrons.  As Matsui and Satz
pointed out more than a decade ago, 
this property makes them excellent candidates as
probes of superdense hadronic matter, created in collisions between
nuclei at high energy \cite{MS85}.  As long as this matter were composed 
of hadrons, so they argued, it would only weakly affect simultaneously 
created heavy quark bound states $(J/\psi,\;\psi',\;\Upsilon,$ etc.).  
On the other hand, if a color deconfined state, i.e. a quark-gluon plasma, 
were formed in the nuclear reaction the formation of heavy quark bound 
states would be severely suppressed, because the attractive color force 
between the quarks is screened by the plasma.  Quantitative calculations
of this effect at finite temperature can be performed in the framework of 
lattice gauge theory.  These calculations yield a state-specific critical 
temperature $T_{\rm d}^{(i)}$ above which the bound state $(i)$ 
dissociates into the plasma.  Whereas the charmonium states dissociate 
soon above the critical temperature $T_{\rm c}$ of the QCD phase transition 
$(T_{\rm d}^{(J/\psi)} \approx 1.2\; T_{\rm c})$, the bottomonium states 
survive up to higher temperatures $(T_{\rm d}^{(\upsilon)} 
\approx 2T_{\rm c})$.

A systematic account of how these characteristic properties of heavy quark 
bound states may be put to use in the study of superdense matter, 
especially for the purpose of establishing the existence of a new 
deconfined phase of QCD, was presented several years ago by Karsch and 
Satz \cite{KS91}.  Since then, new theoretical insight into the elementary 
process of heavy quarkonium production has led to a reassessment of the 
suppression of charmonium production in hadronic interactions with nuclei.  
In fact, as will be discussed in the next section, all experimental data 
taken with nuclear projectiles $(^{16}$O, $^{32}$S) up to 1994 can probably 
be explained as final state interactions of the produced heavy quark pair 
with the nucleons contained in the colliding nuclei.  The discovery of 
enhanced suppression of $J/\psi$ production in collisions between two Pb 
nuclei at the CERN-SPS in 1996, therefore, generated intense interest.  
Does the ``anomalous'' $J/\psi$ suppression observed in Pb + Pb collisions
indicate the formation of a quark-gluon plasma in these reactions?
Has this new state of matter been finally detected?

In order to answer this question, all other conceivable mechanisms
that could explain the observed suppression effect need to be ruled
out.  This is not an easy task because reactions among heavy nuclei at
high energy are processes of great complexity, and there is little
hope that they can be understood in all details.  It is therefore
necessary to allow for a considerable range of parameters in different
reaction models and to exclude competing suppression mechanisms under
a variety of assumed scenarios.

It is the purpose of this review to assess the status of the theoretical 
tools that are needed to establish the possible origin of the observed 
``anomalous'' $J/\psi$ suppression.  The next section summarizes the 
current experimental information on $J/\psi$ and $\psi'$ suppression in 
proton-nucleus and nucleus-nucleus collisions.  The explanation of all 
data (except those from Pb + Pb collisions) in terms of nuclear final 
state interactions will also be discussed in this section.  In Section 3, 
a number of possible mechanisms for the enhanced suppression seen in 
Pb + Pb collision will be surveyed and their current theoretical 
uncertainties will be analyzed.  Section 4 deals with the quantitative 
analysis of the Pb + Pb data in the framework of specific reaction models.  
The review closes with a list of suggestions for future research, mostly 
theoretical, that could help sharpen the tool of charmonium spectroscopy 
for the exploration of the properties of superdense hadronic matter.

The discussion here is mostly based on publications that were available
in June 1997. Later developments are briefly discussed in a separate
section at the end of the review.
For a survey of the state of this research field from a somewhat
different perspective, the reader is urged to also consult the recent
review by Kharzeev \cite{Kha98}.

\section{Phenomenology}

\subsection{Proton-Induced Reactions}

The production of charmonium states in proton interactions with nuclei
has been studied extensively at Fermilab as well as at CERN.  Since
these experiments observe the $\mu^+\mu^-$ decay mode of the $J/\psi$
and $\psi'$, they also provide information on the nuclear target
dependence of the $\mu^+\mu^-$ continuum spectrum produced by light
quark-antiquark annihilation (Drell-Yan process, DY).  Whereas the
Drell-Yan cross section is observed to grow almost exactly in
proportion to the nuclear mass number $A$, the $J/\psi$ and $\psi'$
production cross sections grow less rapidly.  The target dependence is
often parametrized as a power law, $\sigma = \sigma_0 A^{\alpha}$.  A
value $\alpha < 1$ indicates ``suppression'' of charmonium production,
compared with the naive expectation $\sigma/A =$ const. obtained when
one neglects final state interactions.  The argument is that the
strictly linear $A$-dependence of the DY cross section rules out initial
state interactions as the origin of the reduced production of
charmonium states.\footnote{This argument is not strictly correct,
because heavy quarks are predominantly formed by gluon fusion or gluon
scattering.  Gluons might be more susceptible to initial state
interactions than quarks and antiquarks.  There is, indeed, evidence
for enhanced initial state scattering of gluons from the
$p_T$-spectra of heavy quark states produced in $p+\bar p$ reactions,
giving rise to a broader ``intrinsic'' transverse momentum spread of
gluons \cite{Mangano}.}

\begin{figure}[htb]
\vfill
\centerline{
\begin{minipage}[t]{.47\linewidth}\centering
\mbox{\epsfig{file=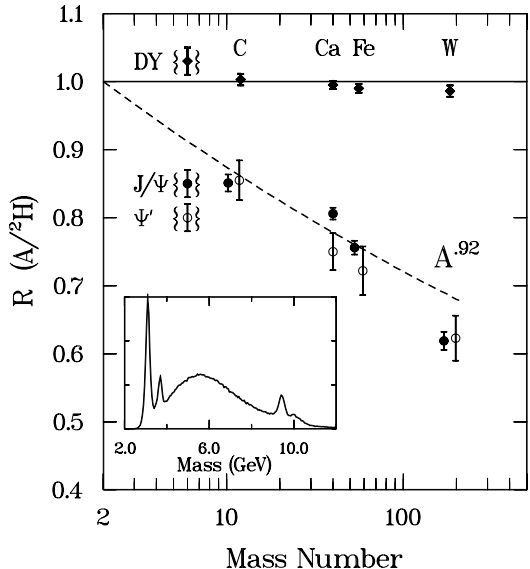,width=.99\linewidth}}
\end{minipage}
\hspace{.06\linewidth}
\begin{minipage}[b]{.47\linewidth}
\caption{Ratio of nuclear cross section compared to deuterium for
Drell-Yan pairs (DY), $J/\psi$ and $\psi'$ production.  Note that the
nuclear suppression for $J/\psi$ and $\psi'$ is the same.  Data from
experiment E772 at Fermilab \protect \cite{E772}.}
\label{fig1}
\end{minipage}}
\end{figure}

The $A$-dependence of the nuclear $J/\psi$ and $\psi'$ cross sections is
remarkably similar.  Both can be fit by the same exponent $\alpha
\approx 0.92$, as shown in Fig.~\ref{fig1}.  Since the mean radius 
of the $\psi'$ is much larger than that of the $J/\psi$, the equally 
strong suppression indicates that the final state interactions occur 
with a state that is not an eigenstate of the $(c\bar c)$ system but 
rather with a common precursor of the $J/\psi$ and $\psi'$.  This does 
not come as a surprise, because the center of mass of the produced 
$(c\bar c)$ pair moves rapidly with respect to the target nucleus.  As 
viewed from the rest frame of the $(c\bar c)$ pair, the nucleus is highly 
Lorentz contracted and thus the $(c\bar c)$ pair leaves the target nucleus 
before the components corresponding to different quantum mechanical 
eigenstates of the charmonium system have had time to decohere.  

The influence of the nuclear target on the decoherence process has been 
studied extensively in different theoretical approaches.  However, the
size of the nuclear suppression remained a great puzzle until quite
recently.  The problem is that the coherent superposition of eigenstates
which is initially produced is characterized by a small geometric size
of order $m_c^{-1} \approx 0.1$ fm.  On the other hand, the sublinear
rise with $A$ of the charmonium production cross sections requires a final 
state interaction cross section $\sigma_{c\bar c N}$ of the order of 
several millibarn.  A comprehensive analysis \cite{KLNS97} of all 
available data on charmonium production in $p+A$ collisions within the 
Glauber model yields a value for the absorption cross section of 
$\sigma_{c\bar c N}^{\rm (abs)} = (7.3 \pm 0.6)$ mb.

\subsection{The Color Octet Model}

The resolution of this apparent paradox can be achieved if one 
assumes that the $(c\bar c)$ pair is originally produced in a color octet
state \cite{KS96} which strongly interacts with its environment.  The
color octet model was originally motivated by unexpectedly large cross
sections for charmonium production observed at the Tevatron.
Perturbative calculations fell short of the measured cross section for 
$J/\psi$ and $\psi'$ production at high $p_T$ by at least one order of 
magnitude.  The observation \cite{Braten} that the data obtained by the 
CDF collaboration can be explained if high-$p_T$ charmonia are 
predominantly formed by gluon fragmentation, led to the hypothesis that 
the charmonium wavefunctions contain a significant component where the 
$(c\bar c)$ pair is in a color octet state accompanied by a soft (valence) 
gluon:
\begin{equation}
\vert\psi(n^3{\rm S}_1)\rangle = O(1) \vert c\bar c [^3{\rm
S}_1^{(1)}]\rangle + O(v_c)\vert c\bar c[^3P_J^{(8)}]g\rangle +
O(v_c^2) \vert c\bar c[^3S_1^{(8)}]gg\rangle + \ldots . \label{e1}
\end{equation}
Here $v_c$ denotes the velocity of the bound charm quarks.  This 
decomposition of the charmonium wavefunction can be unambiguously
defined in the framework of the nonrelativistic expansion of QCD
(NRQCD) developed by Lepage and collaborators \cite{Lepage}.  The higher 
Fock space components are suppressed by powers of the velocity $v_Q$ of 
the heavy quark.  In the heavy quark limit $v_Q\approx {1\over 2}
\alpha_s(m_Q)\ll 1$, producing a well ordered expansion into components 
with an increasing number of valence gluons.  The reason for the 
suppression of higher Fock space components is that ``physical'' gluons 
couple to the color current of the heavy quarks which is proportional to 
their velocity $v_Q=p_Q/m_Q$.  

\begin{figure}[htb]
\vfill
\centerline{
\begin{minipage}[t]{.47\linewidth}\centering
\mbox{\epsfig{file=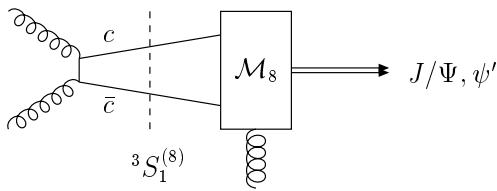,width=.99\linewidth}}
\vspace{0.15truein}
\end{minipage}
\hspace{.06\linewidth}
\begin{minipage}[b]{.47\linewidth}
\caption{Schematic representation of the color-octet formation mechanism
for the charmonium states $J/\psi$ and $\psi'$.  The intermediate state
$^3S_1^{(8)}$ is a color octet which combines with a soft gluon to form
the color-singlet charmonium state.}
\label{fig1a}
\end{minipage}}
\end{figure}

In the limit $\alpha_s\ll 1$, where the structure of the heavy quark
bound states is essentially Coulombic, the momentum scale for the
production of the heavy quark pair, $2m_Q$, is much larger than the
momentum scale associated with the heavy quark bound state,
$p_Q(n^3S_1) = \alpha_s m_Q/2n$.  In the case of the charmonium
ground state, $p_c\approx 1$ GeV compared with $2m_c\approx 3$ GeV,
which still provides for a reasonable separation of scales.  The total
S-matrix element for the production of a $c\bar c[n^3S_1]$ state may
therefore be factorized into a perturbative matrix element for the
elementary production process, such as gluon fusion $(gg\to c\bar c)$ 
or gluon fragmentation ($g^*\to c\bar c$), and a nonperturbative matrix 
element ${\cal M}(c\bar c\to \psi(n^3S_1))$ describing the evolution of 
the elementary $(c\bar c)$ pair into the strong interaction eigenstate 
(1), as illustrated in Fig.~\ref{fig1a}.  
There are different matrix elements, ${\cal M}_1$ and ${\cal M}_8$, 
depending on the color channel in which the $(c\bar c)$ pair is produced.

Assuming that charmonium production at high $p_T$ at the Tevatron is 
dominated by the color octet channel, the matrix elements ${\cal M}_8$ 
for the formation of various eigenstates of the charmonium system can be
deduced empirically \cite{BF,CLW}.  Their values are found to scale in
accordance with the counting rules, the powers of $v_c$, predicted by
(1).  Although the color octet model yields an excellent description
of the measured differential cross sections, one has to caution that
crucial predictions, such as the strong transverse polarization of
$J/\psi$ produced at high $p_T$, have not yet been confirmed.  
It should also be noted that the confirmation of this prediction
would not automatically prove that the color octet mechanism also
dominates $J/\psi$ production at small $p_T$, where the cross section 
is concentrated in nuclear collisions.

An important theoretical problem is how quickly the color octet state
neutralizes its color charge.  Since the neutralization mechanism is,
by assumption, a nonperturbative soft process, it is not rigorously
calculable, but some qualitative estimates are possible.  Since the
force between the $(c\bar c)$ pair in a color octet state is
repulsive, the formation of a bound state requires that the relative
$(c\bar c)$ wavefunction must be a color singlet at distances 
comparable to the average radius of the charmonium state.  More
precisely, the relative magnitude of the octet and singlet components
as a function of separation $r$ should behave as
\begin{equation}
{\vert\psi_{c\bar c}^{(8)}(r)\vert^2\over \vert\psi_{c\bar c}^{(1)}(r)
\vert^2} \sim \exp \left[ -2\int_0^r dr' \sqrt{m_c \vert E_{c\bar
c}-V_8(r')\vert} \right]. \label{e2}
\end{equation}
A crude estimate is obtained by replacing $\vert E_{c\bar c}-V_8\vert$ 
with the binding energy of the charmonium state, $E_B(n^3S_1)$, yielding
the neutralization time
\begin{equation}
\tau_{8\to 1} \approx \left( 2v_c \sqrt{m_c E_B(n^3S_1)}\right)^{-1}
\approx 0.3 n \; {\rm fm}/c. \label{e3}
\end{equation}
Clearly, a more quantitative estimate within the framework of a
coupled-channel potential model for charmonium would be very useful.
In the laboratory frame, the octet-to-singlet transition time is
Lorentz dilated due to the rapid motion of the $(c\bar c)$ pair with
respect to the target nucleus.  Since the average traversed length of
nuclear matter is of the order of 5 fm for a heavy nuclear target, the
precise value of $\tau_{8\to 1}$ becomes important for charmonium
states produced at midrapidity.  However, in first approximation, it is
reasonable to expect that the color neutralization occurs
predominantly outside the target nucleus in $p+A$ experiments
performed at several hundred GeV/$c$ beam energy.

\subsection{Nucleus-Nucleus Collisions}

\begin{figure}[htb]
\vfill
\centerline{
\begin{minipage}[t]{.47\linewidth}\centering
\mbox{\epsfig{file=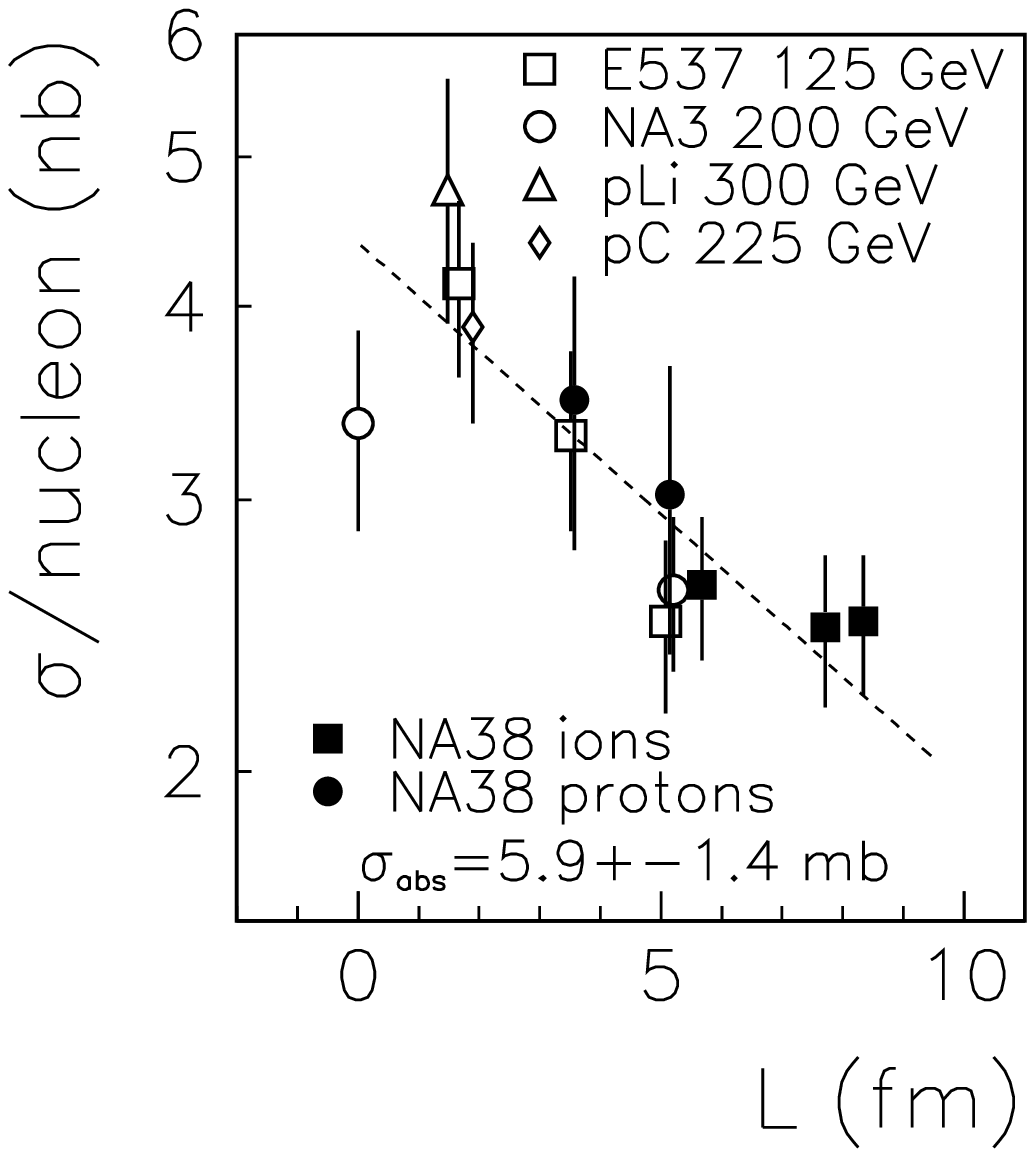,width=.99\linewidth}}
\caption{Effective nuclear absorption cross section in the Glauber
approximation for $J/\psi$ deduced from the NA38 data.  The heavy ion
data (black squares) fall on the same trend as the data from proton
induced reactions.}
\label{fig2}
\end{minipage}
\hspace{.06\linewidth}
\begin{minipage}[t]{.47\linewidth}\centering
\mbox{\epsfig{file=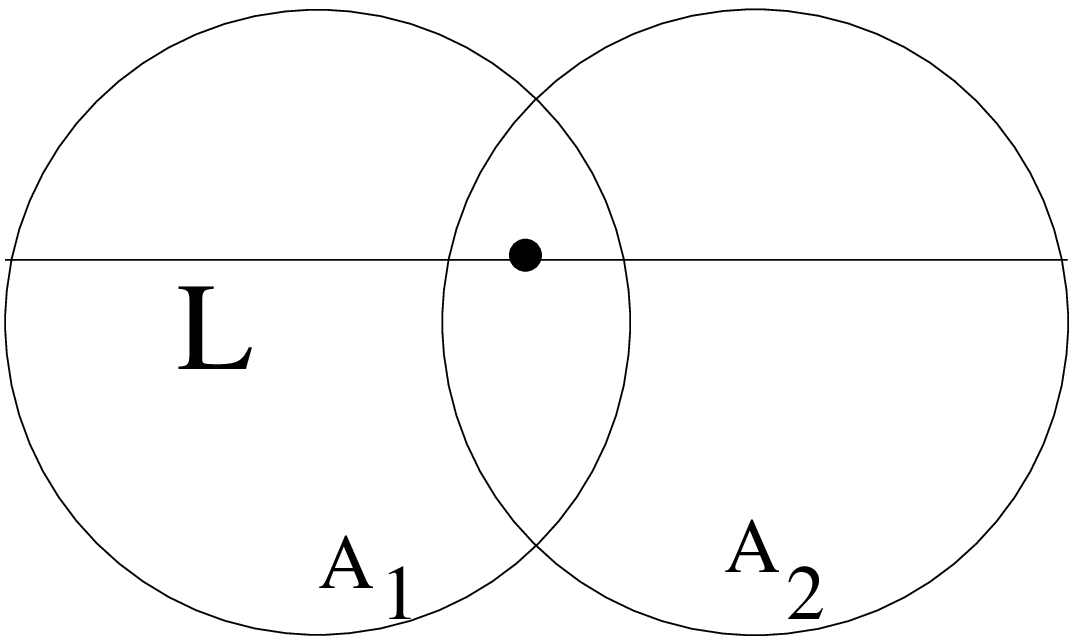,width=.8\linewidth}}
\caption{Illustration of the definition of $L$, the equivalent
thickness of nuclear matter sweeping over the location of the created
$(c\bar c)$ pair.}
\label{fig3}
\end{minipage}}
\end{figure}

The production of $J/\psi$ and $\psi'$ in nucleus-nucleus collisions
has been studied in the experiments NA38 and NA50 at CERN.  NA38
investigated p + W, p + U, $^{16}{\rm O+U}$, and $^{32}{\rm
S+U}$ collisions at 200 GeV/nucleon; NA50 took data for $^{208}$Pb +
Pb collisions at 158 GeV/nucleon.  The results of experiment NA38
showed that the production of $J/\psi$ in $^{16}$O- and $^{32}$S-induced 
reactions follows the nuclear mass dependence expected from $p+A$ 
collisions see Figure (\ref{fig2}). 

The total production cross section $\sigma_{J/\psi}$ grows 
like\footnote{Note that $\rho_NL$ is a Lorentz invariant expression, 
which counts the area density of nucleons interacting with the 
$(c\bar c)$ pair after its formation.}
\begin{equation}
\sigma_{J/\psi} = \sigma_{J/\psi}^{(pp)} \cdot A_1A_2 \exp \left(
-\sigma_{c\bar cN}^{\rm (abs)} \rho_NL (A_1,A_2)\right) \equiv \;
\sigma_{J/\psi}^{(pp)} A_1A_2S_N, \label{e4}
\end{equation}
where $L(A_1,A_2)$ is the average length of nuclear matter traversed
by the $(c\bar c)$ pair after its formation at some moment during the
collision.  The definition of $L$ is illustrated in Fig.~\ref{fig3}.

In the framework of Glauber theory, the nuclear suppression factor $S_N$ 
(and hence $L(A_1,A_2)$) is defined as
\begin{equation}
S_N = \int d^2b\; d^2b'dzdz' \rho_{A_1}(\vec b',z) \rho_{A_2}(\vec
b-\vec b',z') T_{A_1}(z,\vec b') T_{A_2}(z',\vec b-\vec b') \label{e5}
\end{equation}
with the nuclear profile function
\begin{equation}
T_A(z,\vec b) = \exp \left( -(A-1) \sigma_{c\bar cN}^{\rm (abs)}
\int_z^{\infty} dz' \rho_A(\vec b,z')\right). \label{e6}
\end{equation}
Here $\rho_A(\vec b,z)$ is the nuclear density $\sigma_{c\bar
cN}^{\rm (abs)}$ distribution, normalized to unity so that $S_N=1$ for
$p+p$ collisions.

The fact that nuclear production of $J/\psi$ appears to follow the
Glauber model prediction (\ref{e4}) may be interpreted to mean that
nothing ``abnormal'' occurs in nuclear reactions up to projectile mass
$A_1=32$.  This argument was first presented in this form by 
Capella, et al. \cite{Cap88} and by Gerschel and H\"ufner \cite{GH92}.  
It is worth noting that their conclusion
originally was quite controversial because the absorption cross section
$\sigma_{c\bar cN}^{\rm (abs)}$ was expected to be much smaller than the
required 6--7 mb.  This state of affairs has been radically changed by
the introduction of the color-octet model that provides a natural
explanation for $\sigma_{c\bar cN}^{\rm (abs)}$.

\begin{figure}[htb]
\vfill
\centerline{
\begin{minipage}[t]{.47\linewidth}\centering
\mbox{\epsfig{file=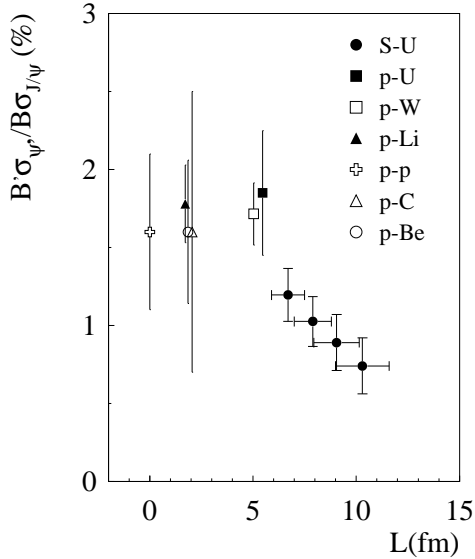,width=.99\linewidth}}
\end{minipage}
\hspace{.06\linewidth}
\begin{minipage}[b]{.47\linewidth}
\caption{The cross section for $\psi'$ production is suppressed relative
to that for $J/\psi$ production in S + U collisions.  The suppression
increases with $L$.  Heavy ion data from experiment NA38.}
\label{fig4}
\end{minipage}}
\end{figure}

The results obtained for $\psi'$ production on the other hand, cannot
be explained by (\ref{e4}).  The $\psi'$ cross section is found to be
more suppressed in $^{32}$S + U reactions than the $J/\psi$
cross section.  This is in marked contrast to $p+A$ collisions where
the ratio $\sigma_{\psi'}/\sigma_{J/\psi}$ is independent of the
target mass.  The different behavior of the $\psi'$ becomes even more
evident when the production of charmonium states is studied as a
function of the total transverse energy $E_T$ carried by the collision
fragments.  Within uncertainties, $E_T$ is a measure of the impact
parameter of the nuclear reaction.  Selected ranges of $E_T$,
therefore, correspond to different values of the average target
thickness $L(A_1,A_2)$ in (\ref{e4}).  The relation between $L$ and
$E_T$ can be modeled in the Glauber approximation or some other
geometric collision model.  The relative suppression of $\psi'$
production compared with $J/\psi$ production increases with growing
$L$ (or $E_T$), as shown in Fig.~\ref{fig4} .  
On the other hand, $\sigma_{J/\psi}$ 
in different $E_T$-windows follows exactly the Glauber model formula
(\ref{e4}), if the $E_T$-dependence of $L$ is taken into account.

The stronger suppression of the $\psi'$ indicates the presence of a
new suppression mechanism in S + U collisions that is absent in $p+A$ 
collisions.  One candidate for this mechanism is absorption by
``comovers'', i.e. by secondary hadrons produced at about the same 
rapidity as the $\psi'$.  Since they are slowly moving with respect to 
the $\psi'$, there is no time dilation of the formation time of the 
$\psi'$ for interactions with these secondary particles.  We will 
return to this issue in section 4, when we discuss models for
comover absorption.

\begin{figure}[htb]
\vfill
\centerline{
\begin{minipage}[t]{.47\linewidth}\centering
\mbox{\epsfig{file=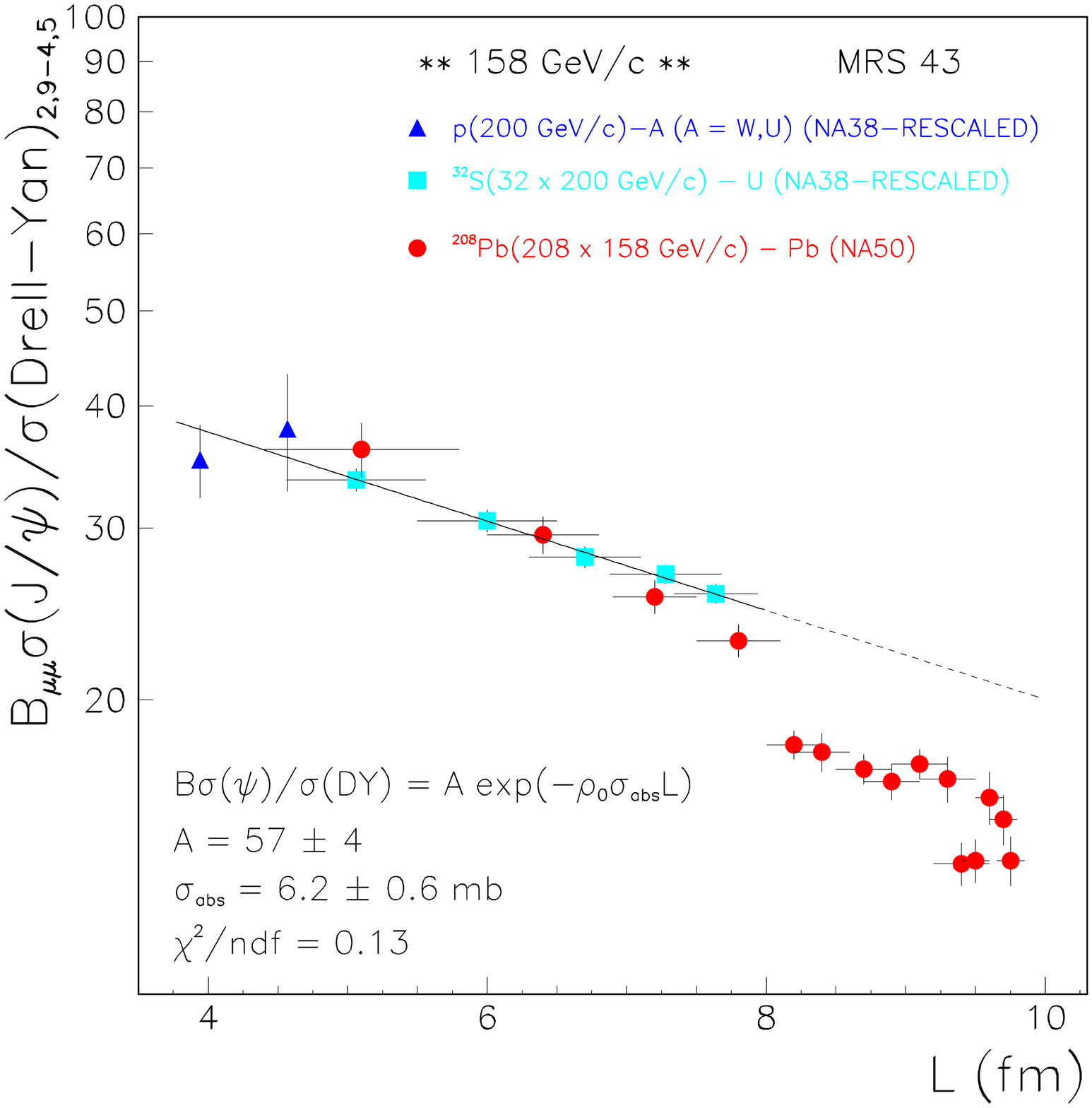,width=.99\linewidth}}
\caption{Results from NA50 for the suppression of $J/\psi$ production
in Pb + Pb collisions as function of the Glauber parameter $L$.
(From \protect\cite{NA50qm}.)}
\label{fig4a}
\end{minipage}
\hspace{.06\linewidth}
\begin{minipage}[t]{.47\linewidth}\centering
\mbox{\epsfig{file=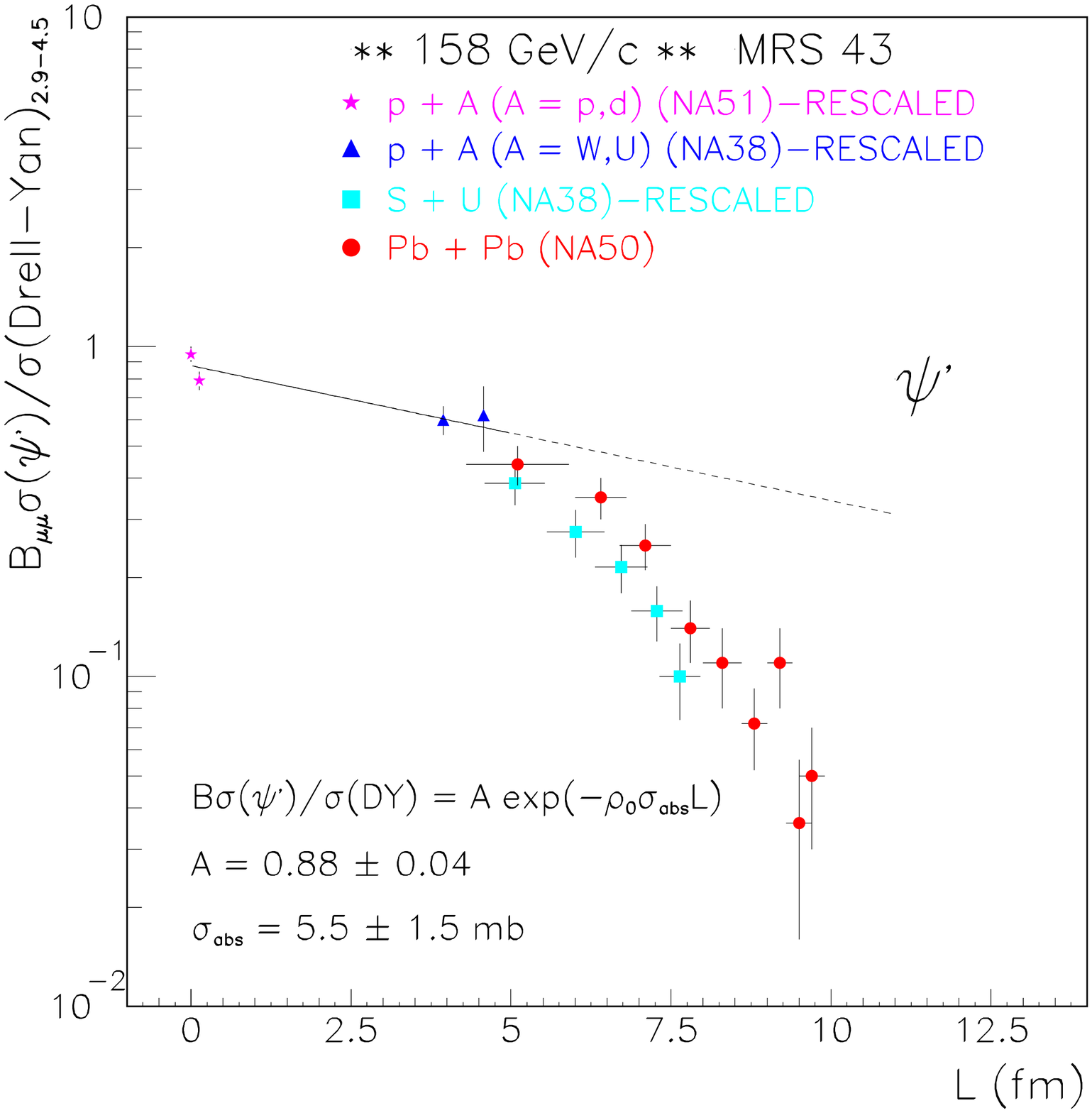,width=.99\linewidth}}
\caption{Results from NA50 for the suppression of $\psi'$ production
in Pb + Pb collisions as function of the Glauber parameter $L$.
(From \protect\cite{NA50qm}.)}
\label{fig4b}
\end{minipage}}
\end{figure}

In Pb + Pb collisions at 158 GeV/nucleon, as shown in Fig.~\ref{fig4a},
also the cross section for $J/\psi$ production is found to be more strongly 
suppressed than predicted by (\ref{e4}). The amount of additional suppression
increases with $L$ (or $E_T$), similar to the effect observed for $\psi'$
production in S + U collisions.  For $\psi'$ production in the
Pb + Pb system, strong additional suppression is observed even in the
lowest $E_T$ range, as shown in Fig.~\ref{fig4b}.  
The suppression factor for the highest $E_T$ (or $L$)
window is only insignificantly larger than that found in the most
central S + U collisions.  Taken together, the S + U and Pb + Pb data 
indicate the onset of a new, ``abnormal'' suppression mechanism for 
charmonium production, first for the $\psi'$ and later for the $J/\psi$, 
which appears to approach saturation for the $\psi'$ in the most central 
Pb + Pb events.

The most obvious difference between $p$ + $A$, S + U, and Pb + Pb
interactions at SPS beam energies is the number of secondary particles. 
As already mentioned before, a significant fraction of the secondary 
hadrons are produced in the same region of rapidity as the charmonium.  
Is it possible to explain the ``abnormal'' suppression as absorption of 
the $J/\psi$ and $\psi'$ on these comoving hadrons?  At first glance, 
this explanation seems eminently plausible.  The absorption cross 
section of the $\psi'$ and comovers should be much larger than that of 
the $J/\psi$, both because of the larger geometrical size of the $\psi'$ 
and its lower energy threshold for dissociation.  Therefore, the comover 
effect should set in much earlier for the $\psi'$ than for the $J/\psi$, 
just as observed.  The question is thus decidedly of a quantitative, 
rather than qualitative, nature:

\begin{enumerate}

\item Is it possible to explain the $A$- and $E_T$-dependence of the
``anomolous'' part of $J/\psi$ and $\psi'$-suppression by a common set
of parameters?

\item Do the required parameters, i.e. comover absorption cross
sections, agree with theoretical estimates or values deduced from other
relevant data?

\end{enumerate}

The following sections address these two questions.  We shall first 
review what theory can tell us about cross sections for the absorption 
of charmonium on hadrons made of light quarks.  We shall then turn to 
phenomenological models that aim at finding a consistent set of 
parameters describing the S + U and Pb + Pb data.

\section{Charmonium Absorption by Comovers}

\subsection{Overview of Mechanisms}

Most mechanisms that have been proposed to describe the absorption of
$J/\psi$ or $\psi'$ on comoving hadronic matter fall into three
categories:\footnote{A comprehensive description of QCD based
theoretical approaches to $J/\psi$ interactions in hadronic matter
can be found in \cite{Khar96}.}

\begin{enumerate}
 \item {\it Deconfinement:} If the comoving matter density is
sufficiently high, the matter may be in the form of a quark-gluon
plasma.  In this case the charmonium states cannot form at all, as
first argued by Matsui and Satz, because the attractive color force
between the $(c\bar c)$ pair is screened by the plasma.  The
color-singlet $(c\bar c)$ potential then has the form
\begin{equation}
V_{cc}(r) = {4\over 3}\alpha_s {e^{-\mu r}\over r}, \label{e7}
\end{equation}
where both $\alpha_s$ and $\mu$ are functions of the density of the
plasma.  For a thermalized plasma with no net baryon excess, finite
temperature perturbation theory predicts
\begin{equation}
\mu^2 = \left( 1+ {1\over 6} N_{\rm f}\right) g^2T^2 \label{e8}
\end{equation}
where $N_{\rm f}=3$ is the number of light quark flavors.  Monte-Carlo
simulations of the SU(3) lattice gauge theory at finite temperature 
\cite{DeG86,KMS88} show that the inverse color screening length $\mu$ in 
the range $T_{\rm c} \le T \le 3T_{\rm c}$ can be parametrized as $\mu 
\approx 2.2T$.  For every bound state of a heavy quark-antiquark 
pair then exists a critical temperature $T_{\rm D}$, above which 
the state disappears.  Detailed studies \cite{KS91}  have revealed that 
$T_{\rm D}(\psi') \approx T_{\rm D}(\chi_c) \approx T_{\rm c}$, 
$T_{\rm D}(J/\psi) \approx 1.2 T_{\rm c}$, and $T_{\rm D}(\Upsilon) 
= 2 T_{\rm c}$.

\item {\it Thermal dissociation:} If the $(c\bar c)$ pair is immersed
in a thermal hadronic gas, its internal modes will eventually also
become thermally occupied.  All those states above the $D\bar D$
threshold will dissociate.   As the temperature rises,
the thermal fraction of dissociated states increases until, eventually, 
the bound states make up a negligible contribution.  In condensed matter 
physics, the temperature above which almost no bound states remain is 
called the Mott temperature.  In this picture, the $J/\psi$ and $\psi'$ 
could essentially disappear even if quarks and gluons remain confined at 
any temperature.  However, the critical question is how fast the internal 
states of a heavy quark pair become thermalized by interactions with a 
hadronic medium.  This is an issue of kinetics, rather than thermodynamics, 
and will be addressed in the next subsection.

\item {\it Pre-thermal dissociation:} In this scenario, charmonium
states could be excited above the dissociation $(D\bar D)$ threshold
by interactions with a medium of comovers that had no time to
equilibrate.  Two popular examples are (i) a pre-thermal parton
plasma \cite{Xu96,Fpp96}, (ii) a system of color flux tubes 
\cite{Neu89,LGM97}.

The first scenario, a pre-equilibrium parton plasma, obviously only
makes sense if the dissociation process is faster than the
thermalization of the plasma:  $\tau_{\rm dis} \le \tau_{\rm eq.}$.
This seems an unlikely case, because parton thermalization is governed
by the cross section $\sigma_{\rm gg}$.  But in the perturbative realm
of QCD, $\sigma_{\rm gg} \approx {9\over 4}\sigma_{\rm gc}$, hence the
plasma should equilibrate faster than a $(c\bar c)$ state can be
dissociated.

Nonetheless, quantitative model studies of the competition between
equilibration of partons and $J/\psi$ dissociation would be useful.
These could be carried out at two levels.  One could study the
dissociation of a fully formed charmonium state by gluon impact in the
environment of a gluon distribution that itself is equilibrating.  Or
one could study the influence of a prethermal gluon bath on the
conversion process from color octet $(c\bar c)$ to the color singlet
charmonium state.  The latter would require some modeling of the
nonperturbative matrix element ${\cal M}_{\rm g}$, but a semi-quantitative
treatment of this mechanism should be possible.

The second scenario, dissociation by color flux tubes, again requires
that the decay rate of flux tubes by light quark pair production,
$\tau_{q\bar q}$, is slower than that of the dissociation rate of the
charmonium state.  This also appears unlikely, because QCD flux tubes
are known to break rapidly $(\tau_{q\bar q} < 1\; {\rm fm}/c)$, while a
$(c\bar c)$ pair cannot be pulled apart too rapidly by virtue of the
large mass of the $c$-quark.  Again, simple model studies of the
competition between these two processes would be easily carried out,
and the dependence on heavy quark mass investigated.  In this scenario, 
too, one can study the possible influence of the presence of coherent 
gluonic fields on the color octet-to-singlet conversion process.

\end{enumerate}

\subsection{Thermal Dissociation Kinetics}

Since, in general, the cross sections $\sigma_{\rm hh}$ among light
hadrons are much larger than those between light hadrons and
charmonium, $\sigma_{{\rm h}\psi}$, it also makes sense to restrict the
study of dissociation of charmonium by light hadrons to a thermal
hadronic environment.\footnote{The validity of this assumption can
be checked with the help of hadronic cascade models.} As we
argued before, the same consideration applies to a partonic
environment above $T_{\rm c}$.  

There are two possible approaches to the dissociation problem:  one at
the quark level where the large mass of the $c$-quark is used to
separate perturbative from nonperturbative aspects of the problem, and
another one that makes use of effective hadronic interactions.  In
this section, we will consider the constituent $(c\bar c)$-approach.
An effective hadronic Lagrangian approach will be discussed in the
following section.

As first explored in detail by Peskin \cite{Pes79} and Bhanot and
Peskin \cite{BP79}, interactions between heavy quark bound states and
light hadrons can be described perturbatively, if the heavy quark mass
$m_Q$ is large enough.  The reason is that the characteristic momentum
scale of the heavy quark bound state is of order $\alpha_{\rm s}m_{\rm
Q}$ which is large.  As a result, only gluons with momenta larger than
$\alpha_{\rm s}m_{\rm Q}$ can resolve the internal color structure of
the $(Q\bar Q)$ bound state and interact with it.  To gluons with much
smaller momenta, the $(Q\bar Q)$ state appears as an inert color
singlet.  The small size of the $(Q\bar Q)$ state then allows for a
systematic multipole expansion of its interaction with external glue
fields, where the dipole interaction dominates at long range.

The potential problem with this approach is that the $c$-quark is not
quite heavy enough.  For instance, the binding energy of the $J/\psi$
is about 650 MeV, which is neither small compared to the reduced mass
${1\over 2}m_c\approx 700$ MeV of the charmonium system, nor large on
the scale of light hadron masses.  Gluons with momenta of order 1
GeV/$c$ or even slightly less can effectively resolve the internal
color structure of the $J/\psi$, but it is not clear that they belong
in the perturbative domain of QCD.

\begin{figure}[htb]
\centerline{\mbox{\epsfig{file=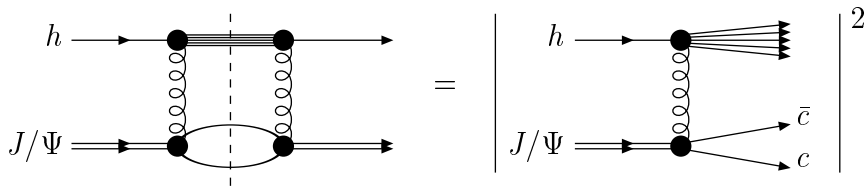,width=.8\linewidth}}}
\caption{Forward scattering between a light hadron and a $J/\psi$ is,
in leading order, described as two-gluon exchange.  The total
absorption cross section is related to the imaginary part of this
forward scattering amplitude.}
\label{fig5}
\end{figure}

By virtue of the optical theorem, the absorptive cross section between
a light hadron $h$ and a charmonium $J/\psi$ can be expressed as the
imaginary part of the forward scattering amplitude where the interaction 
is dominated at long distances by two-gluon exchange (see Figure 
\ref{fig5}).  In this picture, light hadrons interact with the $J/\psi$
only via their glue content.  Kharzeev and Satz \cite{KS94} applied the
Bhanot-Peskin formalism to $\pi-J/\psi$ scattering.  Using the
operator product expansion to separate long-distance physics from the
(assumed) perturbatively calculable structure of the $(c\bar c)$
system, they found that the absorptive cross section is proportional
to the part of the gluon structure function $G_h(x)$ of the
interacting hadron that is sufficiently energetic to dissociate the
$J/\psi$.  Since the gluon structure functions of light hadrons
typically are soft, i.e. $G_h(x) \buildrel x\to 1 \over
\longrightarrow (1-x)^5$, only highly energetic hadrons are capable of
exciting a $J/\psi$ above the dissociation threshold.  The peak cross
section\footnote{The fact that $\sigma_{g\psi}^{\rm (abs)}$ peaks and
then falls off is presumably an artifact of the dipole approximation.  
If higher orders were retained, the cross section should approach a 
constant value at high energy, $\sigma_{g\psi}^{\rm abs} \to 
2\sigma_{gc}$.} for $J/\psi$-dissociation by a gluon in the dipole 
approximation is about 3 mb at a gluon momentum around 1 GeV/$c$.  Due to 
the softness of the gluon structure function, the pion momentum must reach 
5 GeV/$c$ before attaining an absorption cross section in excess of 1 mb.  
The thermally averaged cross section, in this framework, remains less than 
0.1 mb for a pion gas within any realistic temperature range.

\begin{figure}[htb]
\vfill
\centerline{
\begin{minipage}[t]{.47\linewidth}\centering
\mbox{\epsfig{file=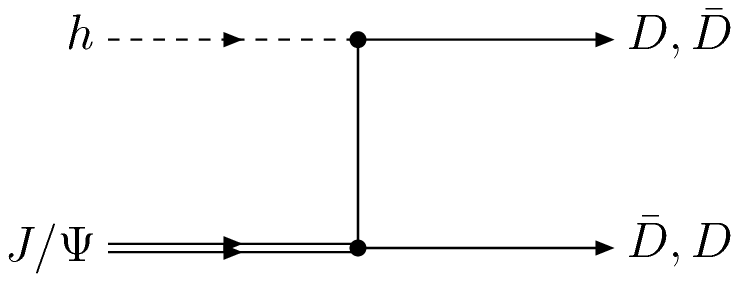,width=.99\linewidth}}
\caption{Inelastic scattering of a light hadron on a $J/\psi$, yielding
two D-mesons in the final state.}
\label{fig6}
\end{minipage}
\hspace{.06\linewidth}
\begin{minipage}[t]{.47\linewidth}\centering
\mbox{\epsfig{file=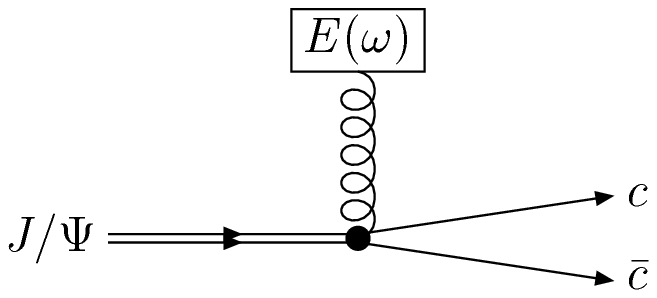,width=.9\linewidth}}
\caption{The $J/\psi$ state is dissociated into a $(c\bar{c})$ octet pair
by absorption of an energetic gluon from its environment.}
\label{fig6a}
\end{minipage}}
\end{figure}

The calculation of \cite{KS94} is clearly incomplete, because pions
are not very efficient at dissociating the $J/\psi$.  Heavier hadrons,
such as the $\rho$-meson, can make significant contributions although 
they are much less abundant than pions.  Moreover, the calculation 
\cite{KS94} does not account for the energy balance between initial and 
final states, i.e. that the total mass of the colliding hadron is available
to convert the $J/\psi$ state into a pair of D-mesons 
(see Fig.~\ref{fig6}).  This effect is especially important for heavier 
mesons, such as the $\rho$.

Repeating the calculation \cite{KS94} in a somewhat more general form,
one finds that the dissociation rate of $J/\psi$-mesons in a (thermal)
hadronic medium is given by (see Appendix A for details)
\begin{equation}
\Gamma_{\rm dis} \approx {1\over\Delta t} \int^{\infty}_{\omega_{\rm
th}} d\omega \langle \vert E(\omega)\vert^2 \rangle\; \pi\alpha_sa^2, 
\label{e11}
\end{equation}
where $a$ is the Bohr radius of the $J/\psi$ and ${1\over\Delta t}
\langle \vert E(\omega)\vert^2\rangle$ denotes the time-average
fluctuation density of color (electric) fields with frequency $\omega$
in the hadronic environment, as illustrated in Fig.~\ref{fig6a}.  
The lower limit of the integral accounts
for the dissociation energy threshold, and the factor $(\alpha_sa^2)$
arises from the coupling to the color dipole moment of the $(c\bar c)$
pair in the $J/\psi$.  Equation (\ref{e11}) is applicable to any kind
of hadronic medium, whether thermal or not.  For example, one could
estimate the color field fluctuations in the framework of a random
field model as proposed by H\"ufner et al. \cite{HGP90}.  For a
thermalized medium, QCD sum rules or lattice gauge theory could be
applied.  Because some heavy quarkonium states, especially the
$\Upsilon$ do not disappear right at the critical temperature, the color
field fluctuation spectral density is also an important quantity to
know in the deconfined phase, where it could be evaluated either in
perturbation theory or by lattice simulations.

\begin{figure}[htb]
\vfill
\centerline{
\begin{minipage}[t]{.40\linewidth}\centering
\mbox{\epsfig{file=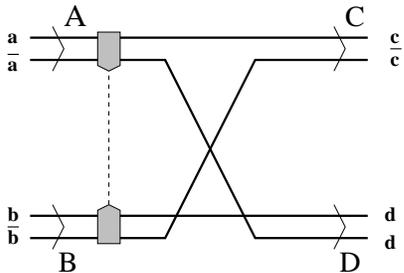,width=.99\linewidth}}
\vspace{0.15truein}
\end{minipage}
\hspace{.06\linewidth}
\begin{minipage}[b]{.54\linewidth}
\caption{In the framework of the constituent quark model, the
absorption of a $J/\psi$ on a light quark meson is viewed as quark
exchange reaction.  At lowest order, four one-gluon exchange diagrams
contribute to the amplitude. (From Martins et al. \protect \cite{MBQ}.)}
\label{fig7}
\end{minipage}}
\end{figure}

The real limitations of (\ref{e11}) lie in the fact that the $J/\psi$
is not truly a perturbative bound state but probes also the confining
part of the $c\bar c$-potential. It is therefore of interest to
investigate the dissociation rate also in the framework of other
approaches that model quark confinement.  One such model is the
constituent quark model of Isgur and Karl \cite{IK79}.  In this model
the reaction
\begin{equation}
\pi + J/\psi \to D^* + \bar D \quad (D+\bar D^*) \label{e12}
\end{equation}
is viewed as quark exchange where a $c$-quark and a light quark change
sides.  The cross section for this reaction was calculated by 
Martins, Blaschke, and Quack \cite{MBQ} in the first Born
approximation (see the diagrammatic representation in Fig.~\ref{fig7}).
Including $D^*\bar D,\; D\bar D^*$, and $D^*\bar D^*$ final states,
the cross section peaks around 1 GeV kinetic energy (c.m.) at a value
of 15 mb (see Fig.~\ref{fig7}).  The magnitude of this cross section 
is solely due to the action of the confining interaction between
the quarks, which is modeled as a ``color-blind'' attractive interaction
between the quarks which has a Gaussian momentum dependence. Because
this interaction is taken as attractive independent of the color
quantum numbers of the affected quark pair, it does not cancel for
the interaction of a light quark with a point-like ($c\bar{c}$) pair
in a color-singlet state, in contrast to the one-gluon exchange
interaction.  Moreover, the magnitude of the obtained cross section 
invalidates the use of the Born approximation, unless a number of 
different partial waves contribute.  This is impossible to ascertain 
from \cite{MBQ}, because no partial wave decomposition of the 
dissociation cross section is presented there.

\begin{figure}[htb]
\vfill
\centerline{
\begin{minipage}[t]{.56\linewidth}\centering
\mbox{\epsfig{file=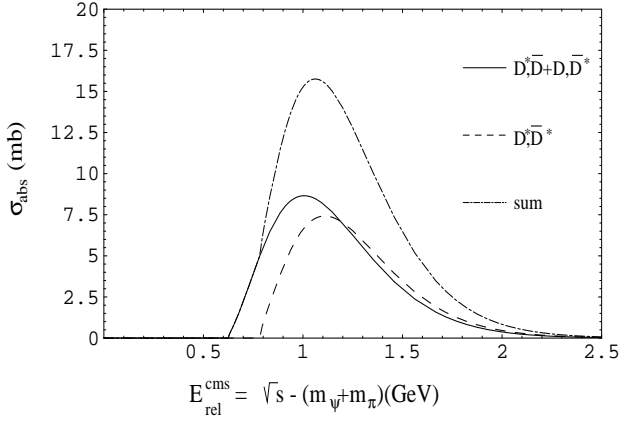,width=.99\linewidth}}
\vspace{0.15truein}
\end{minipage}
\hspace{.06\linewidth}
\begin{minipage}[b]{.38\linewidth}
\caption{Energy dependence of the $J/\psi$ absorption on pions for the
final states $DD^*$ and $D^*D^*$.  (From the preprint version of Martins, 
et al. \protect\cite{MBQ}. Note that the figure in the printed version
differs slightly from the one reproduced here.)}
\label{fig8}
\end{minipage}}
\end{figure}

Since the results obtained within the constituent quark model differ by
orders of magnitude from those obtained in the Bhanot-Peskin approach,
it is useful to consider the absorption process also in an entirely
different framework.  This will be done in the next subsection.

\subsection{Hadronic Dissociation:  Effective Theory}

The most abundant mesons in a hot hadronic gas are $\pi,\; K$, and
$\rho$.  These can induce the following dissociation processes when
encountering a $J/\psi$ particle:
\begin{eqnarray}
&&\pi + J/\psi \to D+ \bar D^*, \quad \bar D + D^*  \label{e21} \\
&&K +J/\psi \to D_s + \bar D^*,\quad \bar D_s +D^*, \quad
D + \bar D_s^*,\quad \bar D+D_s^* \label{e22} \\ 
&&\rho +J/\psi \to D+\bar D, \quad D^*+\bar D^* \label{e23}
\end{eqnarray}
The kinematic thresholds for these reactions are
\begin{center}
(\ref{e21})\quad 640 MeV, \qquad (\ref{e22})\quad 385 MeV, \qquad
(\ref{e23})\quad $-135$ MeV (+155 MeV).
\end{center}
The reaction $\rho+J/\psi\to D+\bar D$ has the lowest total invariant
mass threshold and is, in fact, exothermic.  At thermal equilibrium it
is only suppressed due to the relatively high mass of the
$\rho$-meson which causes $\rho$-mesons to be less abundant than pions 
in a thermal hadron gas.

From a microscopic point of view, all reactions listed above can be
understood as quark exchanges, where the $J/\psi$ transmits a charm
quark to the light meson and picks up a light $(u,\;d,\;{\rm or}\;s)$
quark.  Since similar reactions among light hadrons typically occur with
large cross sections, one can expect that the reactions 
(\ref{e21},\ref{e22},\ref{e23}) also proceed with significant strengths above 
their respective kinematic thresholds. This is, indeed, the result 
obtained by Martins, Blaschke, and Quack \cite{MBQ} which was discussed
at the end of the previous subsection.

\begin{figure}[htb]
\centerline{\mbox{\epsfig{file=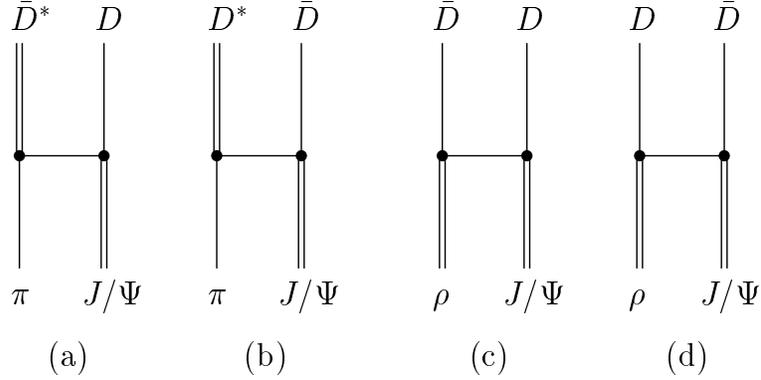,width=.7\linewidth}}}
\caption{Lowest order Feynman diagrams contributing to the charm
exchange reactions \protect (\ref{e21}) and \protect (\ref{e23}).
Single lines represent pseudoscalar mesons; double lines denote
vector mesons.}
\label{fig9}
\end{figure}

Here, we follow a different approach \cite{MM98}.  
In the effective meson theory,
the exchange of a $(c\bar q)$ or $(\bar cq)$ pair, where $q,\bar q$
stands for any light quark, can be described as the exchange of a $D$,
$D_s$, $D^*$, or $D_s^*$ meson between the $J/\psi$ and the incident
light meson (see Fig.~\ref{fig9}).  Near the kinematic threshold, where only
a small number of final states are open, the description in terms of
an effective meson exchange is expected to be valid.

In order to calculate the various Feynman diagrams for the 
reactions (\ref{e21},\ref{e22},\ref{e23}) we need to construct the effective 
three-meson vertices.  We do so by invoking a strongly broken U(4)
flavor symmetry with the vector mesons playing the role of quasi-gauge 
bosons.  Denoting the 16-plet of pseudoscalar mesons $(\pi,\eta,\eta',
K,D,D_s,\eta_c)$ by $\Phi=\phi_iT_i$, where the $T_i$ are the $U(4)$ 
generators, and the vector meson 16-plet $(\omega,\rho,\phi,K^*,D^*,D_s^*,
\psi)$ by ${\cal V}^{\mu} = V_i^{\mu}T_i$, the free meson Lagrangian reads
\begin{equation}
{\cal L}_0 = {\rm tr} (\partial^{\mu}\Phi^{\dagger}\;
\partial_{\mu}\Phi) - {\rm tr} (\partial^{\mu}{\cal
V}^{\dagger,\nu})(\partial_{\mu}{\cal V}_{\nu}-\partial_{\nu}{\cal
V}_{\mu}) - {\rm tr} (\Phi^{\dagger}M_{\rm P}\Phi) + {1\over 2}
{\rm tr}({\cal V}^{\mu_{\dagger}}M_{\rm V}{\cal V}_{\mu}). \label{e13}
\end{equation}
Here $M_{\rm P}$ and $M_{\rm V}$ denote the mass matrices for the 
pseudoscalar and vector mesons, respectively.  Because of the heavy mass 
of the charm quark, $M_{\rm P}$ and $M_{\rm V}$ break the U(4) symmetry 
strongly down to U(3), the mass of the strange quark introduces a weaker 
breaking to U(2), and the axial anomaly further breaks the U(2) symmetry 
to SU(2) in the case of the pseudoscalar mesons.  All these symmetry
breakings are embodied in the physical mass matrices.  It is
convenient, in the following, to work with the mass eigenstates.

The meson couplings are obtained by replacing the space-time
derivatives $\partial_{\mu}$ by the ``gauge covariant'' derivatives
\begin{equation}
D_{\mu} = \partial_{\mu} - ig {\cal V}_{\mu}. \label{e14}
\end{equation}
In first order in the coupling constant $g$ this procedure leads to
the following interactions
\begin{eqnarray}
{\cal L}_{\rm int} &= &ig\; {\rm tr} \left( \Phi^{\dagger} {\cal
V}^{\mu \dagger}\partial_{\mu}\Phi - \partial^{\mu}\Phi^{\dagger}{\cal
V}_{\mu}\Phi\right) \nonumber \\
&&\quad + ig\; {\rm tr} \left( \partial^{\mu}{\cal V}^{\dagger\nu}
[{\cal V}_{\mu},{\cal V}_{\nu}] - [{\cal V}^{\dagger \mu},{\cal
V}^{\dagger \nu}]\partial_{\mu}{\cal V}_{\nu}\right). \label{e15}
\end{eqnarray}
If the U(4) flavor symmetry were exact, we would expect all
couplings given by the same constant $g$.  In view of the significant
breaking of the flavor symmetry we anticipate that the effective
coupling constants for different 3-meson vertices will have different
values.  We will see below to what extent this is true.

In order to describe $\pi$- and $\rho$- induced $J/\psi$ dissociation,
we need the following vertices:  $\psi DD,\;\psi D^*D^*,\;
\pi DD^*,\;\rho DD$, and $\rho D^* D^*$.  From (\ref{e15}) we derive
the following interactions:
\begin{eqnarray}
{\cal L}_{\psi_{DD}} &= &ig_{\psi_{DD}} \psi^{\mu}\left( \bar
D\partial_{\mu}D - (\partial_{\mu}\bar D) D\right) \nonumber \\
{\cal L}_{\psi D^*D^*} &= &-ig_{\psi D^*D^*}\psi^{\mu} \left( \bar
D^{*\nu}\partial_{\mu}D_{\nu}^* - (\partial_{\mu}\bar D^{*\nu})
D_{\nu}^*\right) \nonumber \\
{\cal L}_{\pi DD^*} &= &{i\over 2}g_{\pi DD^*} \left( \bar D \tau_i
D^{*\mu}\partial_{\mu}\pi_i - \partial^{\mu} D \tau_i D_{\mu}^*
\pi_i - \partial_{\mu}\pi_i \bar D^{*\mu} \tau_i D + \pi_i \bar D^{*\mu} 
\tau_i \partial_{\mu}D \right) \nonumber \\
{\cal L}_{\rho DD} &= &{i\over 2}g_{\rho DD} \rho_i^{\mu} \left( \bar D
\tau_i \partial_{\mu} D - \partial_{\mu}\bar D \tau_i D \right) \nonumber \\
{\cal L}_{\rho D^*D^*} &= &-{i\over 2} g_{\rho D^*D^*} \rho_i^{\mu} 
\left( \bar D^{*\nu} \tau_i \partial_{\mu} D_{\nu}^* - (\partial_{\mu} 
\bar D^{*\nu}) \tau_i D_{\nu}^* \right). \label{e16}
\end{eqnarray}
The coupling constants $g_{\psi DD},\; g_{\psi D^*D^*},\; g_{\rho DD}$
and $g_{\rho D^*D^*}$ can be derived from the $D$ and $D^*$ electric
form factors in the framework of the vector meson dominance (VMD) model.

If $\gamma_{\rm V}$ denotes the photon-vector meson ${\cal V}$ mixing
amplitude, the standard VMD analysis \cite{VMD}  yields the relations
\begin{equation}
\gamma_{\rho}f_{\rho} = em_{\rho}^2, \qquad \gamma_{\psi} f_{\psi}
= {2\over 3}em_{\psi}^2 \label{e17}
\end{equation}
where universality of the vector meson couplings is assumed:
\begin{equation}
f_{\rho} = g_{\rho DD} = g_{\rho D^*D^*}, \qquad f_{\psi} = g_{\psi DD} 
= g_{\psi D^*D^*}. \label{e18}
\end{equation}
The photon mixing amplitudes $\gamma_{\rm V}$ can be determined from the 
leptonic vector meson decay widths \cite{VMD}:
\begin{equation}
\Gamma_{\rm Vee} = {1\over 3} \alpha {\gamma_{\rm V}^2\over m_{\rm V}^3}.
\label{e19}
\end{equation}
Inserting the experimental numbers, we find
\begin{equation}
f_{\rho} \approx 5.6, \qquad f_{\psi} \approx 7.7. \label{e20}
\end{equation}
As is well known, different ways of deriving the value of the
``universal'' $\rho$-meson coupling $f_{\rho}$ yield values differing by
about 20\%.  The coupling constants (\ref{e20}) must therefore be
considered to be given with an error of this order of magnitude.

Finally, the $\pi$-meson coupling between $D$ and $D^*$ can be
obtained from the decay width of the $D^*$-meson:  $D^*\to D\pi$.
Unfortunately, only an upper bound for this decay rate is known at
present, corresponding to $g_{\pi DD^*}<15$.  Theoretical
estimates for this decay rate, based on QCD sum rules \cite{Bel95}, yield
\begin{equation}
g_{\pi DD^*} \approx 8.8.
\end{equation}
We adopt this value here.

At the hadronic level, the charm exchange reaction between the
$J/\psi$ and a light hadron can proceed either by exchange of a $D$-
or a $D^*$-meson.  Here we shall only consider the $D$-exchange
reactions, because they have an acceptable high-energy behavior even
in the absence of form factors for the vertices.  The $D^*$-exchange
reactions have cross sections that rise rapidly with energy due to the
exchange of longitudinally polarized $D^*$-mesons.  This growth
characteristic of the exchange of massive vector bosons is obviously
unphysical and will be cut off by the mesonic form factors long
before it becomes significant.  However, this would require that we
introduce an unknown parameter (the characteristic momentum scale of
the internal mesonic structure).  In order to avoid this complication,
we here neglect $D^*$-exchange cross sections.  One final remark
on $D^*$-exchange:  Regge theory dictates that, in the high energy
limit, the charm exchange reaction is dominated by the exchange of the
$D^*$-trajectory.  However, here we are not interested in charm
exchange at high energies but near the kinematical threshold, because
the relative motion of the hadrons is limited to thermal momenta.  

\begin{figure}[htb]
\vfill
\centerline{
\begin{minipage}[t]{.47\linewidth}\centering
\mbox{\epsfig{file=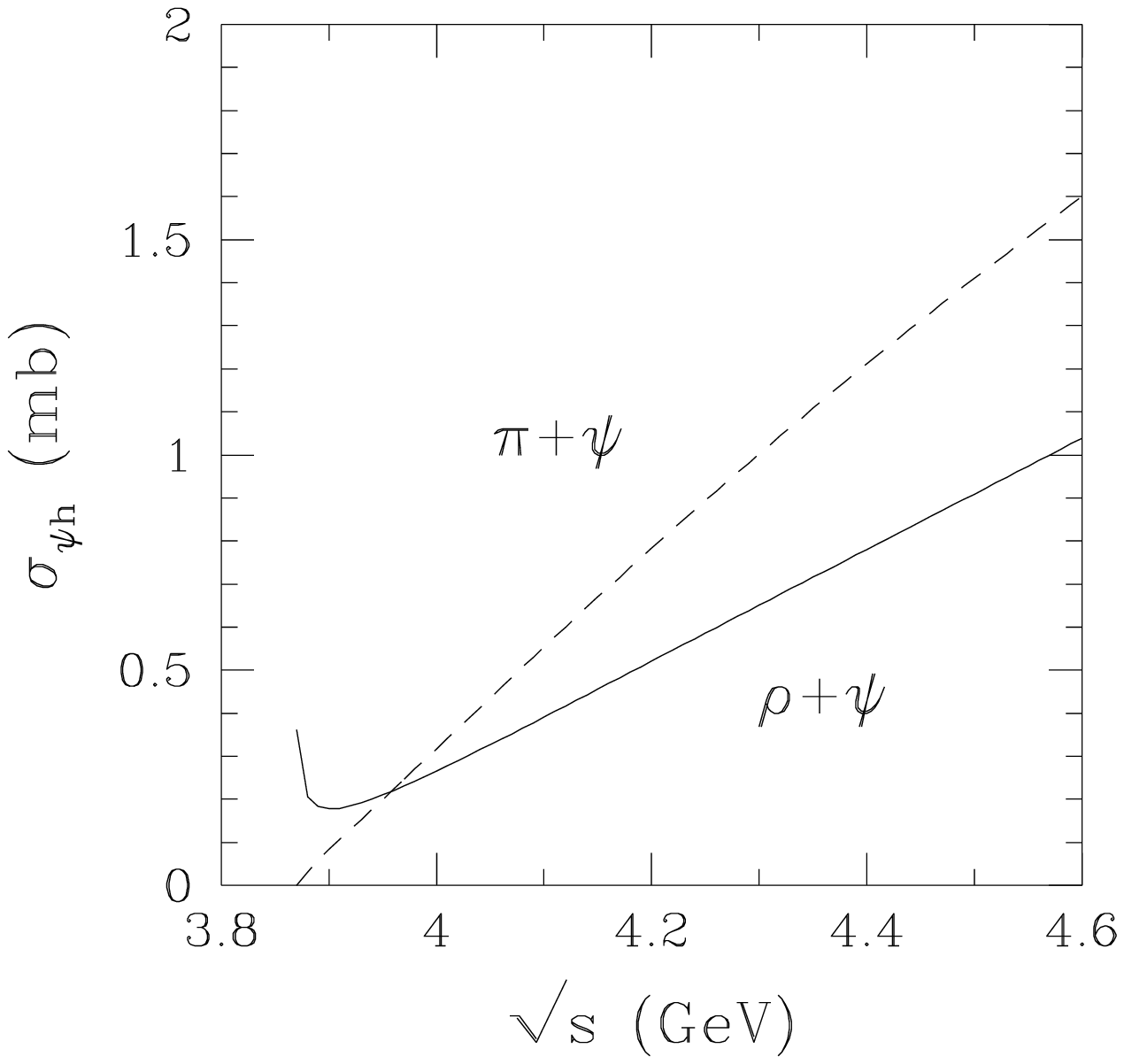,width=.99\linewidth}}
\caption{Cross sections for the charm exchange reactions described by
the diagrams of Figure \protect \ref{fig9}, as functions of c.m.
energy. Dashed line: pions, solid line: $\rho$-mesons.}
\label{fig10}
\end{minipage}
\hspace{.06\linewidth}
\begin{minipage}[t]{.47\linewidth}\centering
\mbox{\epsfig{file=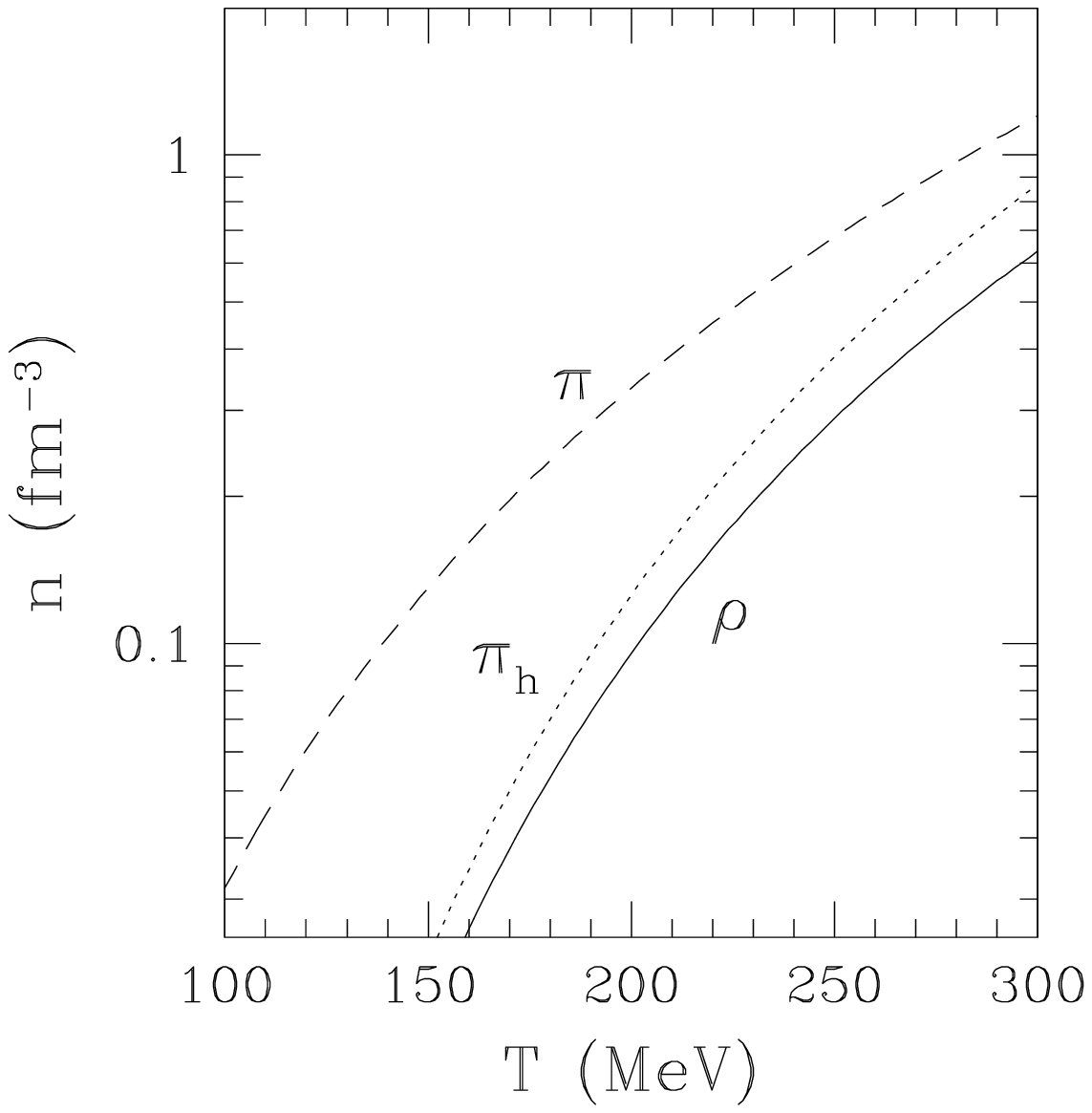,width=.99\linewidth}}
\caption{Thermal pion and $\rho$-meson densities in an ideal hadron gas
as function of temperature. Dashed line: pions, solid line: $\rho$-mesons.
The dotted line represents the density of pions above the dissociation
threshold for $J/\psi$.}
\label{fig11}
\end{minipage}}
\end{figure}

Analytical expressions for the amplitudes and cross sections for the
processes corresponding to the Feynman diagrams in Fig.~\ref{fig9} are
given in Appendix B.  The charm exchange cross sections are plotted in
Fig.~\ref{fig10} as functions of the center-of-mass energy
$\sqrt{s}$.  The $\pi +J/\psi$ cross section (dashed line) starts at
zero, because the reaction is endothermic, whereas the $\rho+ J/\psi$
cross section (solid line) is finite at the threshold because of its
exothermic nature.  Of course, the reaction rate vanishes at threshold
in both cases.  Note that the cross sections for these two $J/\psi$ 
absorption reactions are of similar magnitude over the energy
range relevant to a thermal meson environment.

\begin{figure}[htb]
\vfill
\centerline{
\begin{minipage}[t]{.47\linewidth}\centering
\mbox{\epsfig{file=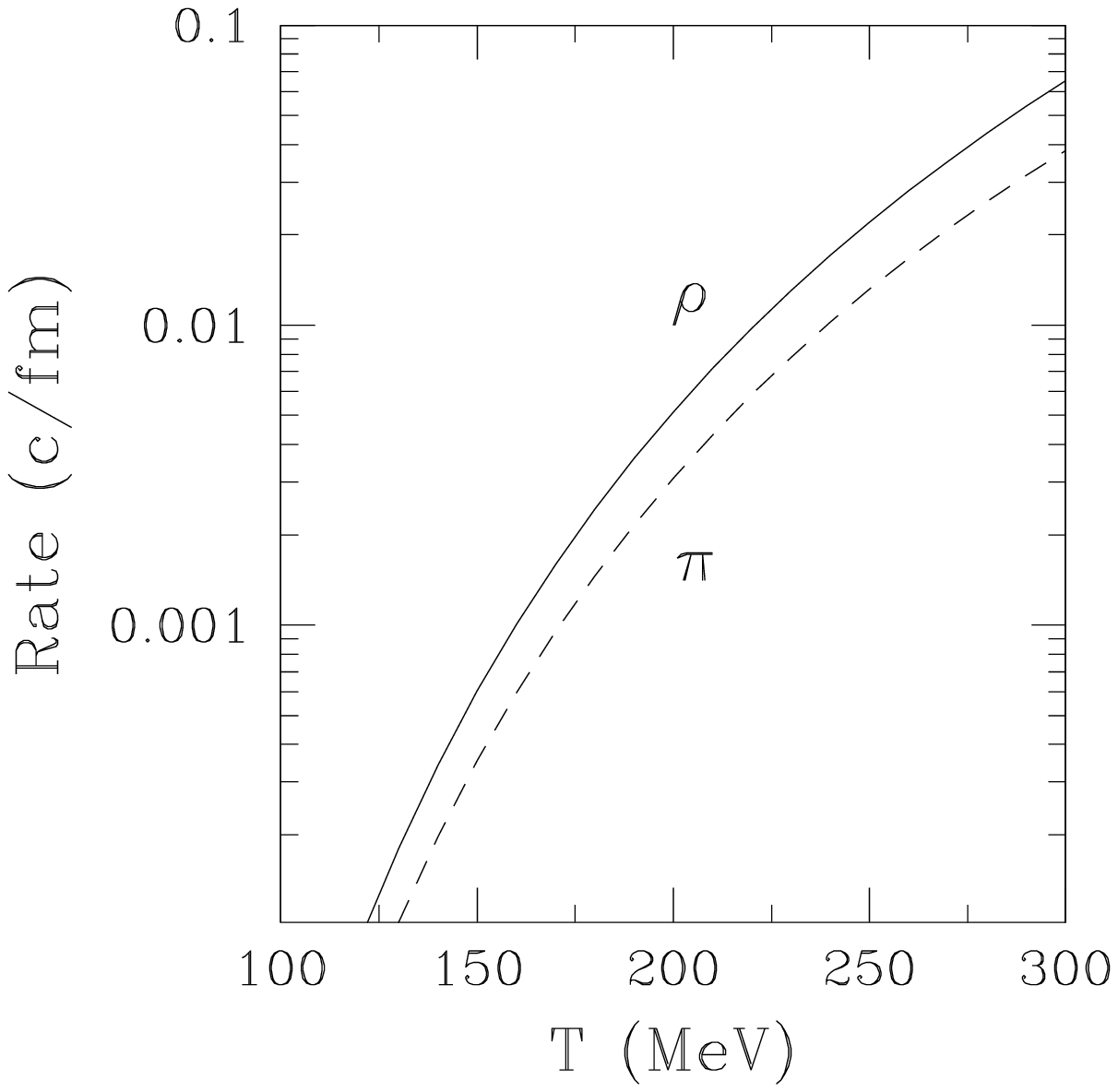,width=.99\linewidth}}
\caption{Thermal $J/\psi$ absorption rates as function of temperature.
The pion and $\rho$-meson rates are shown separately.}
\label{fig12}
\end{minipage}
\hspace{.06\linewidth}
\begin{minipage}[t]{.47\linewidth}\centering
\mbox{\epsfig{file=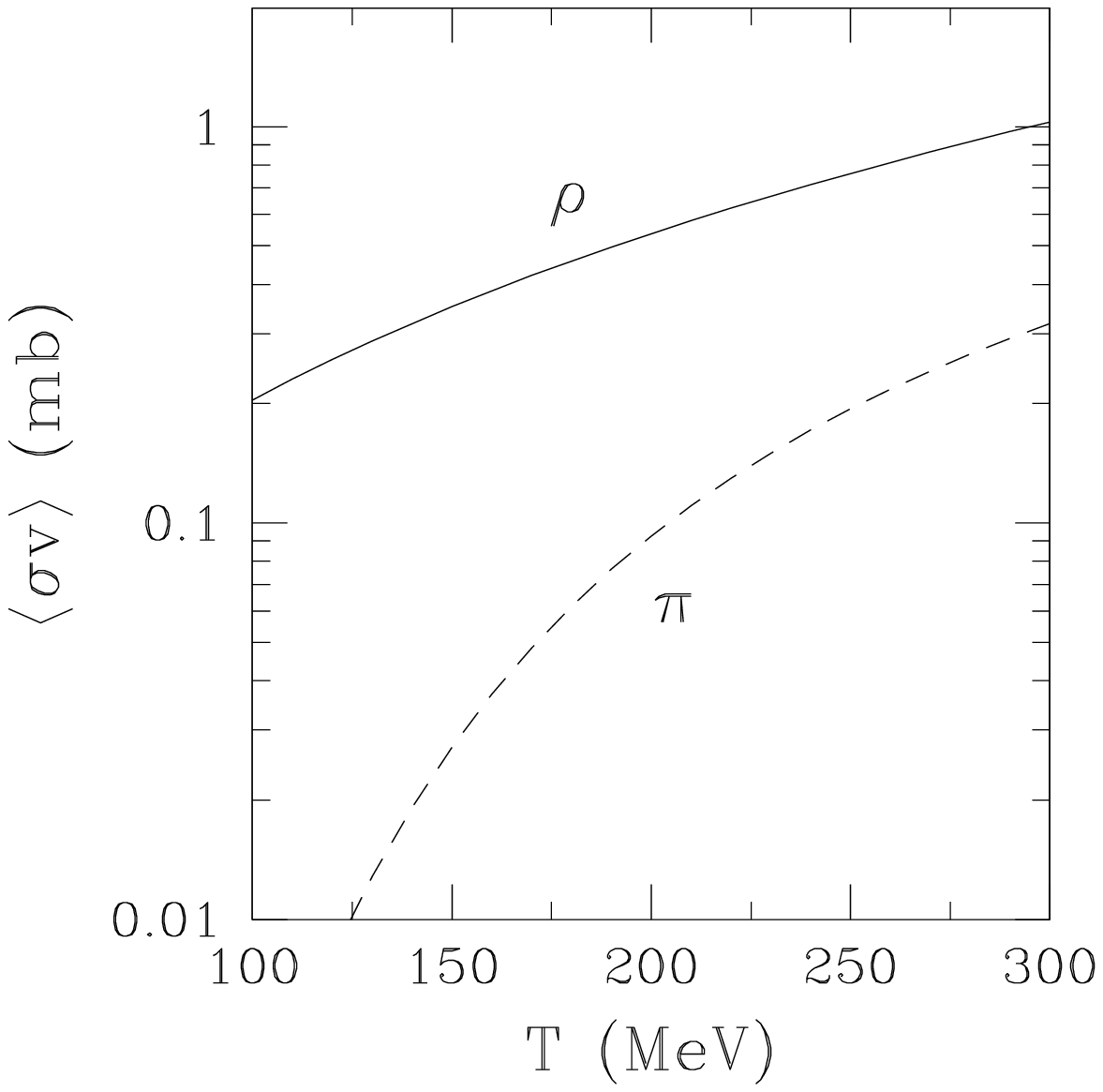,width=.99\linewidth}}
\caption{Thermally averaged $J/\psi$ absorption cross sections as function 
of temperature.  The pion and $\rho$-meson rates are shown separately.}
\label{fig12a}
\end{minipage}}
\end{figure}

Figure \ref{fig11} shows the pion and $\rho$-meson densities in an ideal
thermal meson gas as a function of temperature.  In the temperature
range where a hadronic gas is likely to exist, $T \le 170$ MeV, pions
are far more abundant than $\rho$-mesons, but most of these pions do
not have sufficient energy to initiate the dissociation of a
$J/\psi$.  In fact, if one counts only those pions and $\rho$-mesons
above the kinematical threshold for $J/\psi$-dissociation, $\pi$- and
$\rho$-mesons are about equally abundant, or rather rare, because the
effective density does not exceed 0.1 fm$^{-3}$ even at $T = 200$ MeV, 
where the hadron picture of a thermal environment probably already fails.
Figure \ref{fig12} shows the $J/\psi$ absorption rates in a thermal meson 
gas, as a function of temperature.  Even at the (unrealistically) high
temperature $T = 300$ MeV, the thermal dissociation rate is still so 
small that it corresponds to a lifetime around 10 fm/$c$.  Thermal 
dissociation at $T \le$ 200 MeV is  completely negligible on the time 
scale of the lifetime of a hot hadronic gas state in nuclear collisions.

One can ask the question whether a form factor should be included
in the meson vertices used to evaluate the Feynman diagrams in
Fig.~\ref{fig10}. In principle, the concept of vector meson
dominance assumes that the off-shell behavior of the vector meson 
propagators describes the form factor of the pseudoscalar mesons
correctly, when the pseudoscalar mesons are on-shell. In the diagrams
of Fig.~\ref{fig10}, however, the exchanged $D$-meson is off-shell
by a considerable amount, whereas the vector mesons all
appear on-shell either in the initial or final state. It is unclear
whether an additional form factor is needed in this situation. In
any case, our result represents an upper limit, because an additional
form factor would reduce the charm exchange rates. (E.g. a Gaussian
form factor $\exp(-Q^2/Q_0^2)$ with $Q_0 = 1.5$ GeV would further reduce 
the absorption rates by slightly more than one order of magnitude.)

As we noted above, our analysis also remains incomplete because of the
neglect of $D^*$-exchange reactions.  It would be most interesting to
consider also these reactions in the framework of a complete effective
Lagrangian describing the interactions of pseudoscalar and vector
mesons with quark content $(q\bar q),\; (Q\bar q),\; (q\bar Q)$, and
$(Q\bar Q)$, where $q$ stands for any light quark flavor and $Q$
denotes a heavy quark.  Such a Lagrangian embodying chiral symmetry for 
the light quarks has recently been given by Chan \cite{Chan}. Loop
corrections to the tree diagrams would also permit the study of form
factor effects in this approach.

\section{Models versus the Data}

\subsection{Glauber Theory}

Probably the most conservative approach to calculate nuclear effects
of suppression of charmonium production is based on the general
framework of Glauber theory.  This was, in fact, the starting point of
the early analysis of Gerschel and H\"ufner \cite{GH92}, and it forms
the basis of more recent analyses of the available $p+A$ and $A+A$
data \cite{KLNS97,AC97}.  In the Glauber formalism the suppression
factor $S_N$ due to absorption on nucleons is given by (\ref{e5}).
The effect of absorption by comoving secondary hadrons is included by
multiplying the integrand in (\ref{e5}) by an additional exponential
absorption factor:
\begin{equation}
S = \int d^2b\; d^2b'\; {d S_N\over d^2b\; d^2b'}\; T_{\rm co}(b,b') 
\end{equation}
with
\begin{equation}
T_{\rm co}(b,b') = \exp{\left[-\sigma_{\rm co}^{\rm eff} 
\int_{\tau_0}^{\tau_f}\; 
{dN_{\rm co}(b,b') \over dy} d\tau\right]}. \label{24}
\end{equation}
Here $dN_{\rm co}(b,b')/dy$ is the comover density per unit of
rapidity at impact parameter $\vec b,\vec b'$ with respect to the two
nuclear centers, and $\sigma_{\rm co}^{\rm eff} = \langle \sigma_{{\rm
h}\psi}^{\rm (abs)}v\rangle$ is the effective $J/\psi$ absorption cross 
section by comovers.  The effect of quark deconfinement (complete 
absorption) can be easily described in this framework by setting 
$\sigma_{\rm co}^{\rm eff} = \infty$ for $dN_{\rm co}/dy > 
dN_{\rm co}^{\rm crit}/dy$.  $\tau_0$ corresponds to the formation time 
of the dense comover gas (typically taken as $\tau_0 = 1-2\; {\rm fm}/c)$, 
and $\tau_f$ denotes the time of hadronic freeze-out.

\begin{figure}[htb]
\centerline{\mbox{\epsfig{file=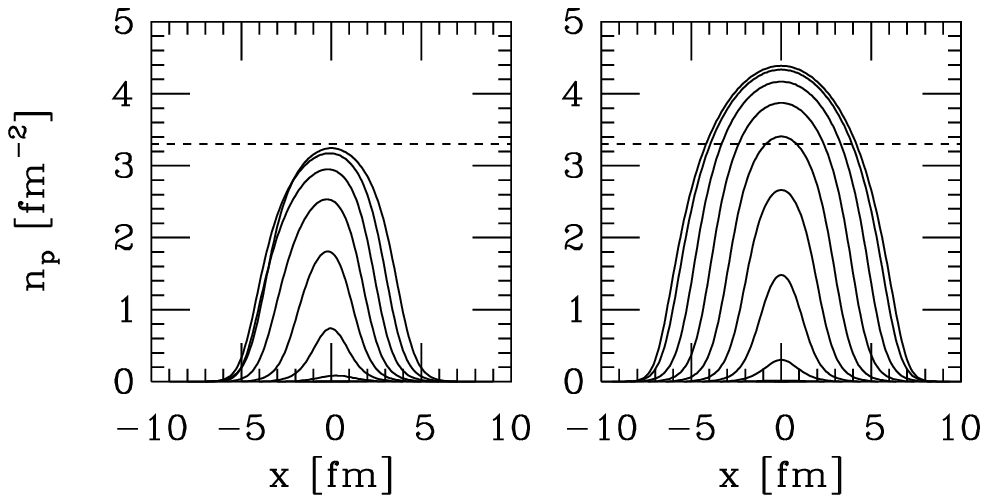,width=.85\linewidth}}}
\caption{Initial energy densities according to the Bjorken formula
\protect (\ref{Bf}) are shown as functions of impact
parameter for (left) S + U and (right) Pb + Pb collisions. 
(From ref.~\protect\cite{BO96}.)}
\label{fig13}
\end{figure}

It is reasonable to assume that the comover density is proportional to
the transverse energy $dE_T/dy$ produced in a nucleus-nucleus
collision at impact parameter $b$.  A detailed analysis of
$dE_T/dy$ measurements in S + U and Pb + Pb collisions \cite{KLNS97}
shows that they can be well described by assuming that the transverse
energy production is proportional to the number of participant
projectile and target nucleons, as calculated in a geometrical picture
or within the Glauber model itself.  This is in agreement with a large
body of data indicating that the ``wounded nucleon'' model \cite{BBG76}
provides a good description of global observables in nuclear
collisions.  Using Bjorken's formula for the comoving energy density
\cite{Bj83}
\begin{equation}
\epsilon_0 = {dE_T/dy\over \pi R^2\tau_0}, \label{Bf}
\end{equation}
where $\pi R^2$ denotes the effective nuclear geometric cross section, and 
setting $\tau_0 = 1\; {\rm fm}/c$, the initial energy densities for 
different impact parameters in Pb + Pb and S + U collisions are found as shown
in Fig.~\ref{fig13}.  Clearly, for most impact parameters the
density achieved in Pb + Pb collisions exceeds that produced in even the 
most central S + U collisions.  This observation opens the door \cite{BO96}
for an explanation of the anomalous $J/\psi$ production observed in
Pb + Pb collisions as the effect of deconfinement if a critical energy
density $\epsilon_{\rm c} \approx 2.75\; {\rm GeV/fm}^3$ is exceeded (more
precisely, if $\epsilon_0\tau_0$ exceeds 2.75 GeV/fm$^2c$).

The earliest systematical analysis of suppression by comovers is due
to Gavin and Vogt \cite{GV90}.  Using the parameters 
$\sigma_{c\bar cN}^{\rm (abs)}= 4.8\; {\rm mb}$, 
$\sigma_{\rm co}^{\rm eff} = 3.2$ mb, and $\tau_0$ = 2 fm/$c$ that were 
originally derived from the S + U data \cite{NA38} they 
find a good overall agreement with the measured suppression in Pb + Pb
collisions \cite{GV97}.  However, it must be noted that, according to our 
present understanding, their value of $\sigma_{c\bar cN}^{\rm (abs)}$ may 
be too small to consistently explain the $p\; + A$ data \cite{KLNS97}, and 
$\sigma_{\rm co}^{\rm eff}$ is most likely much too large.  The point 
here is that at $\tau_0$ = 2 fm/$c$ the formation of a color singlet 
$(c\bar c)$ state should be completed and, hence, the comover absorption 
should be of hadronic size $(\sigma_{\rm co}^{\rm eff} < 1$ mb) as 
discussed in the previous section.

\begin{figure}[htb]
\vfill
\centerline{
\begin{minipage}[t]{.47\linewidth}\centering
\mbox{\epsfig{file=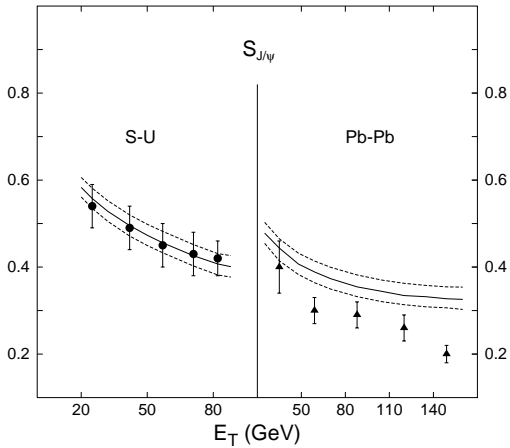,width=.99\linewidth}}
\vspace{-0.1truein}
\end{minipage}
\hspace{.06\linewidth}
\begin{minipage}[b]{.47\linewidth}\centering
\caption{$J/\psi$ suppression factor due to absorption on nucleons
as obtained in the Glauber approximation with an absorption cross
section of $7.3 \pm 0.6$ mb, in comparison with the NA38/50 data
for the S + U and Pb + Pb systems (from \protect\cite{KLNS97}). The
suppression in S + U is nicely explained, but the Glauber model fails
in the Pb + Pb system. The excellent agreement for S + U appears to leave 
no room for comover absorption in this system.}
\label{fig14}
\end{minipage}}
\end{figure}

On the basis of their Glauber model analysis of $p$ + $A$ data on $J/\psi$
suppression, Kharzeev et al. \cite{KLNS97} argue that there is no room for 
absorption by comovers in S + U collisions (left part of Fig.~\ref{fig14}).  
In order to explain the significant additional suppression 
observed in the Pb + Pb data (right part of Fig.~\ref{fig14}), these 
authors must invoke a suppression mechanism that sets in abruptly and 
strongly if $dE_T/dy$, or rather the area density of wounded nucleons, 
exceeds a certain critical value.

This is also found in the analysis of the same data by Armesto and
Capella \cite{AC97}, who obtain acceptable fits with the parameter
choices $\sigma_{c\bar cN}^{\rm (abs)}= 6.7 (7.3)$ mb, 
$\sigma_{\rm co}^{\rm eff} = 0.55 (1.0)$ mb, and critical area density 
of wounded nucleons of $dN^{\rm crit}/dy = 1.15 (2.5)\; {\rm fm}^{-2}$.

What these two analyses demonstrate is that the present body of data
on $J/\psi$ production in $p$ + $A$ collisions leave very little room for
additional suppression effects in S + U collisions, at least within
the framework of the Glauber model.  In order to obtain sufficiently
strong suppression effects in Pb + Pb, a strongly nonlinear dependence
of the exponent of (\ref{24}) on the comover density (here modeled as
a threshold effect) is then required.  Whereas Kharzeev et
al.\cite{KLNS97} argue that this indicates a qualitatively new mechanism 
(such as color deconfinement), the authors of \cite{AC97} do not reach 
this conclusion.

\begin{figure}[htb]
\centerline{\mbox{\epsfig{file=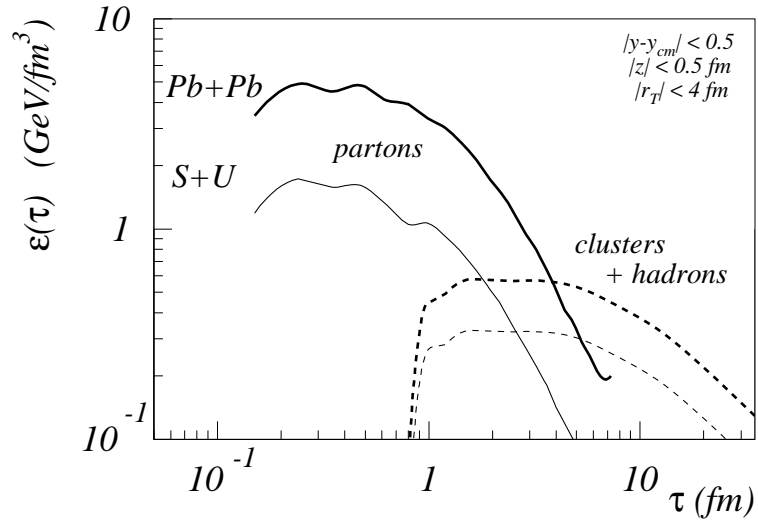,width=.7\linewidth}}}
\caption{Partonic energy densities for S + U and Pb + Pb in the parton
cascade model \protect \cite{GS97}.}
\label{fig15}
\end{figure}

The results for the hadronic absorption rate described above arguably point
toward a failure of the Glauber model to describe the Pb + Pb data, at
least if only hadronic interactions are considered.\footnote{Not all
authors share this opinion. In particular, Gavin\cite{Gav96} and 
Capella\cite{AC97} have maintained that absorption by hadronic or
``pre-hadronic'' comovers can describe all observations.}  Whether the
reason for this failure is the emergence of a new suppression
mechanism, or whether the $A$-dependence of the effective comover
density is strongly underestimated by the Glauber approach, is
impossible to tell from the existing data.  The latter possibility has
been recently addressed in \cite{GM97}, where it was pointed out that
the comovers responsible for $J/\psi$ suppression have to be
``semihard'' in order to resolve the color dipole structure of the
$J/\psi$.  A perturbative QCD picture then suggests that the area
density of these comovers grows as $(A_1,A_2)^{1/3}$ in collisions at
the SPS, much faster than the $(A_1^{1/3}+A_2^{1/3})$ growth predicted
by the wounded nucleon model.  Of course, such comovers would be
better represented as partonic excitations, not hadronic ones.
Indeed, a hybrid parton cascade model \cite{GS97} predicts just such a
dependence for the partonic components of secondary particles (see
Fig.~\ref{fig15}).

\begin{figure}[htb]
\vfill
\centerline{
\begin{minipage}[t]{.47\linewidth}\centering
\mbox{\epsfig{file=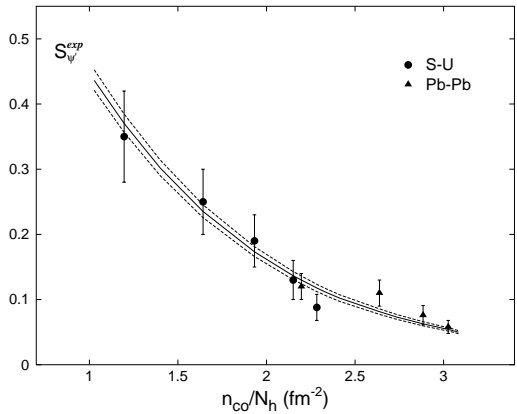,width=.99\linewidth}}
\vspace{-0.1truein}
\end{minipage}
\hspace{.06\linewidth}
\begin{minipage}[b]{.47\linewidth}\centering
\caption{$\psi'$ suppression factor due to absorption on nucleons
(with a cross section of $7.3 \pm 0.6$ mb) as well as absorption on comoving
hadrons (with a cross section of about 10 mb), in comparison with the 
NA38/50 data for the S + U and Pb + Pb systems (from \protect\cite{KLNS97}). 
Comover absorption is required in both systems if one wants to obtain
agreement between the data and the Glauber model.}
\label{fig15a}
\end{minipage}}
\end{figure}

On the other hand, the suppression of the $\psi'$ state is so strong
already in S + U collisions that its description requires comover 
suppression.  This is almost trivially obvious from the fact that the
$\psi'/(J/\psi)$ ratio is constant in $p$ + $A$ collisions, but drops
significantly and progressively with decreasing impact parameter
(growing $dE_T/dy$) in S + U \cite{Bag95}.  This has been analyzed by 
many authors \cite{KLNS97,GV90,Wong96,CKKG96} who all reach similar 
conclusions (see Fig.~\ref{fig15a} for an example).  
Values for the absorption of the $\psi'$ by comovers are 
obtained in the range $\sigma_{\rm co}^{\rm eff}(\psi') = 8$ mb, which 
is quite reasonable in view of the rather large size of the $\psi'$ and 
of its low dissociation threshold.  In fact, the authors of \cite{CKKG96} 
argue that the $\psi'$ suppression observed in Pb + Pb is not quite as 
strong as expected on the basis of the S + U data, and that a regeneration
mechanism must be introduced to obtain a good fit of the Pb + Pb data.

\subsection{Microscopic Models}

Microscopic models of heavy ion reactions allow us to study the
question whether the models based on Glauber theory adequately
describe the space-time dynamics of these reactions.  This includes
questions such as whether the time dependence of the comover density
and the comover composition is modeled correctly in the Glauber approach 
and whether the transverse expansion, which is neglected in the
Glauber model, plays a significant role.

An extensive study of such questions was recently made by Bratkovskaya
and Cassing in the framework of a relativistic hadronic cascade model
\cite{BC97}.  The model treats the nuclear collision as a sequence of
binary hadronic scatterings with the possibility of a formation time for
all newly created particles.  When a $J/\psi$ particle is produced in a
primary nucleon-nucleon collision, it can interact with another hadron
and be dissociated.  The authors assume that the $(c\bar c)$ pair is
originally produced in the color octet state and converts to a color
singlet $J/\psi$ after $\tau_f = 0.7$ fm/$c$.

It is useful to begin taking a look at the general space-time structure
of a nuclear collision event.  The Pb + Pb system differs from higher
systems (p + Pb, S + U) in that a significant fraction of $J/\psi$ are
formed as color singlets during the time period in which the primary
nucleon-nucleon collisions occur.  This is so because at the CERN-SPS
energy, corresponding to $\gamma_{\rm cm} \approx 9$, the Pb nuclei are
Lorentz contracted to a longitudinal width of only about 1.5 fm,
exceeding the assumed $J/\psi$ formation length by a factor two.
Accordingly, the $J/\psi$ states are on average produced in a more
violent environment than in the case for lighter collision systems.  The
density of comovers is higher in Pb + Pb than in S + U and falls off
more slowly.

The Cassing-Bratkovskaya model predicts a peak density of pions
which slightly exceeds \break 1 fm$^3$ in the Pb + Pb system.  A look at our
Fig.~\ref{fig11} shows that this pion density is equivalent to a 
temperature $T\approx 300$ MeV under equilibrium conditions.  The same 
holds true for the peak density of $\rho$-mesons (about 0.6 fm$^{-3}$), 
indicating that the comover abundances may be close to chemical, if not 
thermal, equilibrium.  In other words, the model predicts hadronic 
comover densities far in excess of those in the range of the presumed 
validity of the hadronic phase.  The peak densities predicted for the 
S + U system are lower and correspond to those of a thermal environment of
``only'' $T\approx 250$ MeV.

\begin{figure}[htb]
\vfill
\centerline{
\begin{minipage}[b]{.47\linewidth}\centering
\mbox{\epsfig{file=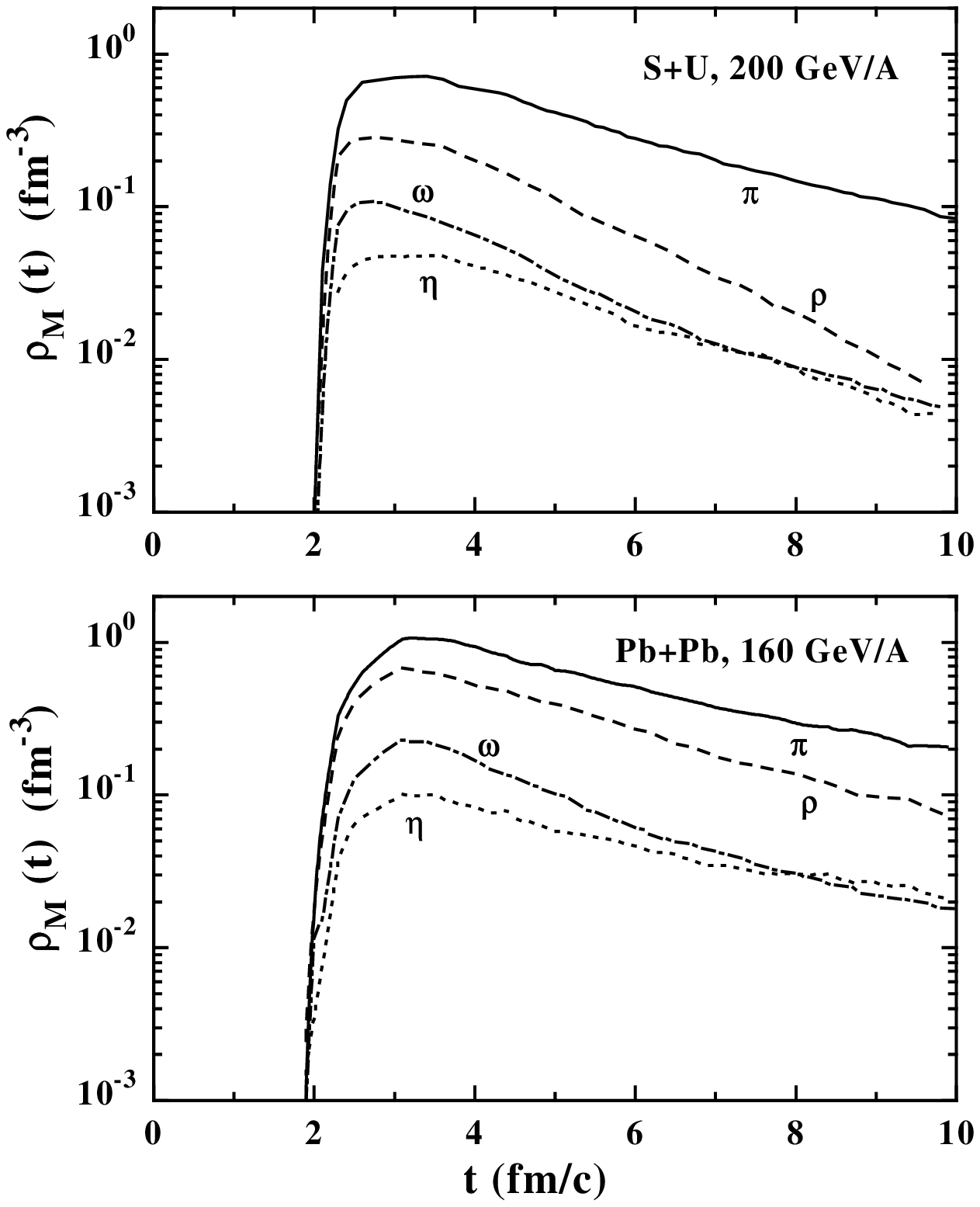,width=.99\linewidth}}
\caption{Density of comovers as function of time in the cascade model
of Cassing and Bratkovskaya.}
\label{fig16}
\end{minipage}
\hspace{.06\linewidth}
\begin{minipage}[b]{.47\linewidth}\centering
\mbox{\epsfig{file=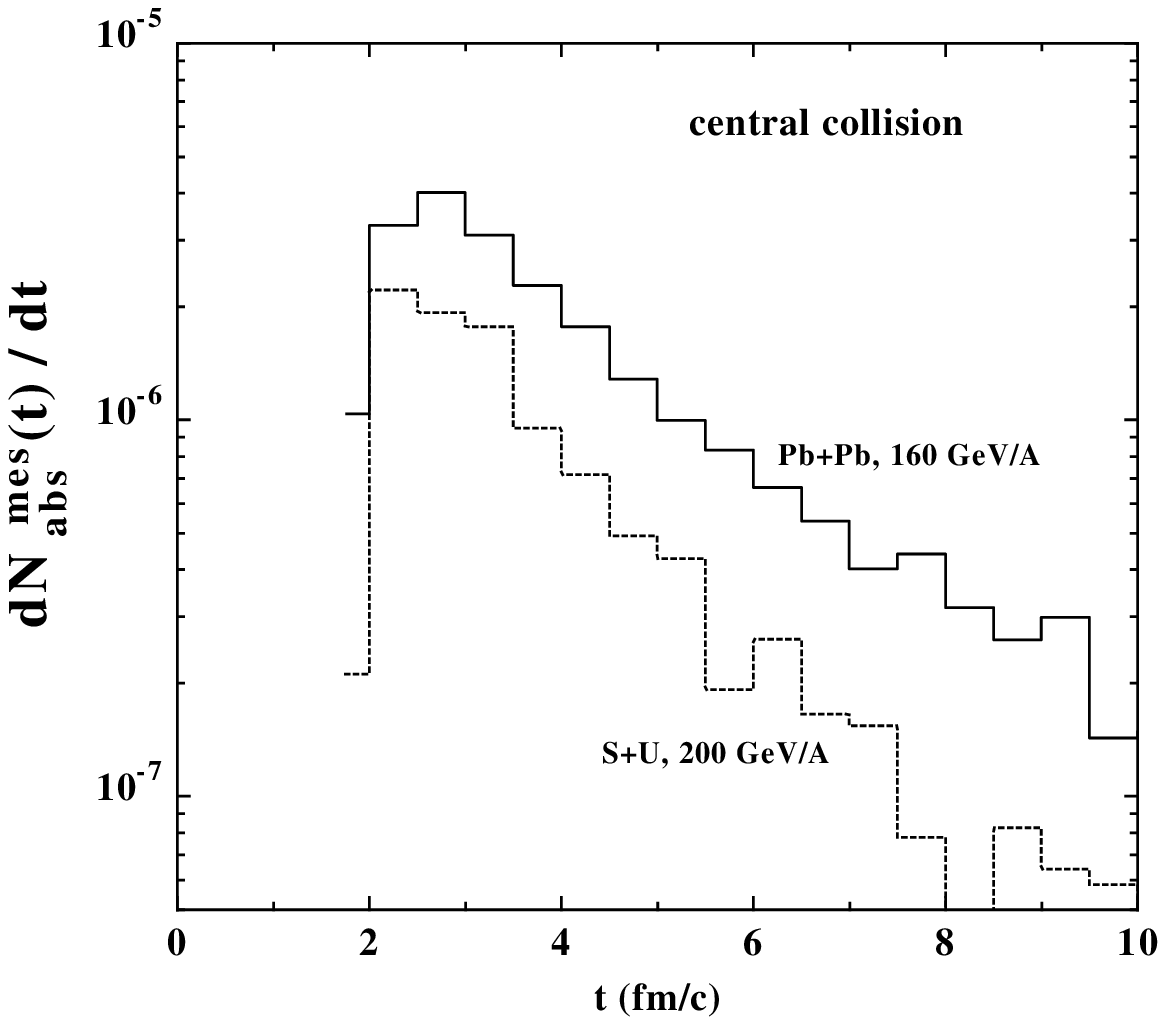,width=.99\linewidth}}
\caption{Time distribution of $J/\psi$ absorption events.  Most
absorptions occur within 3 fm/$c$ after the onset of the reaction.}
\label{fig17}
\end{minipage}}
\end{figure}

Figure \ref{fig17} shows that most of the absorption of $J/\psi$ by 
comovers in this model occurs within the first 3 fm/$c$ after the
initial reaction.  At that time $(t = 5$ fm/$c$ in Fig.~\ref{fig17}), 
the comover density still corresponds to a thermal bath at $T\approx 250$ 
MeV in Pb + Pb and $T \approx 210$ MeV in S + U.  This confirms the 
conclusion reached in Section III.C, that significant dissociation rates 
for $J/\psi$ in a thermal hadronic environment require temperatures far 
beyond $T_c$, at least in the range $T\approx 250-300$ MeV.  

A second interesting observation is that the time interval during which
the $J/\psi$ can be efficiently dissociated is so short that transverse
expansion effects can be safely neglected.  Hence, the Glauber
approximation should be an excellent approximation, and in many respects
preferable to microscopic cascade models, because it allows for a more
consistent and economical treatment of quantum effects.\footnote{Cascade
models can be useful tools for estimating quantities such as the
effective optical thickness of the comoving matter, $\int d\tau v
\sigma_{\rm abs} dN(\tau)/dy$, as input into simplified Glauber
calculations.  However, one needs to exercise caution since quantum
coherence effects that may affect $N(\tau)$ are neglected.}

In conclusion, hadronic cascade models provide useful tools for testing
the consistency of the hadronic comover suppression scenario.  At
present, their results confirm that this scenario has difficulty
explaining an observed ``anomalous'' suppression within the range of
validity of the hadronic gas model.

\section{Addendum: Developments since June 1997}

Since mid-1997, a large number of articles have appeared addressing
the implications of the new data from experiment NA50 on $J/\psi$ 
and $\psi'$ production in Pb + Pb collisions. This activity was
triggered by the new results presented by the NA50 collaboration,
first at the Quark Matter 1997 conference in Tsukuba \cite{NA50qm}
and later at the Moriond meeting in March 1998. The new data from
the analysis of the 1996 Pb beam run at CERN, reproduced in Figures
\ref{fig4a} and \ref{fig4b}, revealed a pronounced sudden drop in 
the ratio of $J/\psi$ production to Drell-Yan pairs in the
Pb + Pb system at an impact parameter around 8 fm (corresponding to
a nuclear absorption length $L \approx 8$ fm). Furthermore, the
new data confirmed the presence of an ``anomalous'' suppression
effect in Pb + Pb, and their much higher statistics leaves hardly
any doubt that the effect is real.

After a careful reanalysis of the viability of hadronic comover models 
Vogt \cite{Vogt97} concluded that the data for p + A and S + U
collisions leave insufficient room for hadronic comovers to make
an explanation of the suppression of $J/\psi$ in Pb + Pb in terms
of hadronic absorption viable. This agrees with the results of
Kharzeev, et al. \cite{KLNS97}. Vogt finds no difficulty with an
interpretation of the $\psi'$ data from NA50 in terms of hadronic
absorption. Vogt's analysis includes the suppression effects on the
feed-down to $J/\psi$ from the excited states $\chi_c$ (normally 
about 30\%) and $\psi'$ (12\%). The S + U data can be fitted with
$\sigma^{\rm (abs)}_{c\bar{c}N}=7.3$ mb and no comover suppression
or with $\sigma^{\rm (abs)}_{c\bar{c}N}=4.8$ mb and 
$\sigma_{\rm co}=0.67$ mb, 1.6 mb, and 2.5 mb for $J/\psi$, $\chi_c$,
and $\psi'$, respectively. The extrapolation to Pb + Pb then fails
to explain the NA50 data, if the comover density is scaled as $E_T$, 
as expected on the basis of multiparticle production models,
such as the wounded nucleon model. A reasonable agreement with the
Pb + Pb data would require a much faster than linear rise of the 
comover density with $E_T$, such as $E_T^{5/3}$.\footnote{Such a
fast rise could be explained by assuming that the comover density
relevant for $J/\psi$ absorption scales like other hard QCD processes,
as suggested by the arguments presented in \cite{GM97}.}

Conclusions similar to those of Vogt are reached by Wong \cite{Wong97},
who analyzes in detail some implications of the competing color
singlet and octet production mechanisms for the $J/\psi$ and $\psi'$.
Martins and Blaschke \cite{MB98}, based on their hadronic charm
exchange model \cite{MBQ}, also rule out the hadronic absorption scenario.

On the other hand, Dias de Deus and Seixas \cite{Dias98} argue that
a very simple, schematic model of comover suppression fits the
NA50 data in a roundabout way; however, their model does not explain
the apparently different behavior of the low- and high-$E_T$ regions
in the Pb + Pb system. Frankel and Frati \cite{FF97} argue that the
observed suppression can be explained in terms of the energy loss
of partons (gluons) as they travel through the colliding nuclei.
Apart from being unable to describe a sudden drop in the $J/\psi$
yield, it is not entirely clear how this model can escape the
stringent limits on energy loss set by the Drell-Yan phenomenology.

A number of authors have quantitatively investigated the effectiveness
of various absorption mechanisms for $J/\psi$ in dense matter. The
reaction $\pi+J/\psi \to \psi'+\pi$ in the presence of a thermal
pion gas was studied in two articles.  Chen and Savage \cite{CS98} 
analyzed the reaction within the framework of chiral perturbation 
theory, finding a very small average cross section of order 0.01 mb 
at thermal energies.  Sorge, Shuryak, and Zahed \cite{SSZ97}
made the assumption that the $\pi\pi$ spectrum in the decay $\psi'
\to J/\psi\pi\pi$ is dominated by a scalar resonance. In their model
the ratio $\psi'/(J/\psi)$ can chemically equilibrate to the value
0.05 observed in the highest $E_T$ bins of the Pb + Pb and S + U
data, if the mass of the scalar resonance drops at high density,
enhancing the reaction rate.
Shuryak and Teaney \cite{ST98} calculated the reaction $\pi+J/\psi
\to \eta_c+\rho$, but found that the thermal rate was only about 
$10^{-3}$ (fm/$c$)$^{-1}$.

Several authors \cite{Vogt97,MB98,KNS97} have emphasized that the 
major part, if not all, of the additional suppression of the $J/\psi$ 
yield seen in the Pb + Pb data can be understood as virtually complete
absorption of the fraction (about 30\%) of $J/\psi$ which originates 
from decays of $\chi_c$.  The deconfinement scenario favors this
explanation, because the $\chi_c$ disappears almost immediately at
$T_c$, while the $J/\psi$ survives to higher temperatures. This
picture is supported by recent, improved lattice calculations
\cite{Fing97}. QCD-based calculations of the absorption cross sections 
of $J/\psi$ and $\chi_c$ on nucleons also show the higher vulnerability
of the $\chi_c$ state at low energies \cite{KS95a}. Gerland et al.
\cite{Ger98} argue that the stronger absorption of $\chi_c$ is already
needed to explain the p + A data and hence must play a significant
role in any explanation of the heavy ion data for $J/\psi$ suppression.

The picture of $J/\psi$ suppression by QCD strings has been embraced
with considerable enthusiasm lately. Following \cite{LGM97},
Geiss et al. \cite{Geiss} simulated this effect within the HSD
model discussed in section 4.2.  They found reasonable
agreement with the NA50 data by setting the string radius to 0.2--0.3 fm
and assuming that any $J/\psi$ immediately dissociates when entering
a string.  With a string area density of about 3/fm$^2$, this
implies that at least half the transverse cross section of the reaction
volume in Pb + Pb is filled with QCD strings. A related, but more
schematic study by Nardi and Satz \cite{NS98} concludes that the
sudden drop in the $J/\psi$ yield observed by NA50 coincides with
the point where QCD strings become so dense that large overlapping
clusters develop for an assumed string radius of 0.2 fm. In this picture
the drop is associated with a color percolation phase transition which
causes the disappearance of the $\chi_c$ state.  The existence of
such a phase transition was also emphasized by Braun, Pajares,
and Ranft \cite{BPR97}, who studied the percolation of strings 
analytically as well as by simulations.  Much earlier studies of
$J/\psi$ absorption reached similar conclusions \cite{Neu89}.

\section{Summary and Outlook}

Have heavy quark bound states, the $J/\psi$ and $\psi'$, fulfilled their
promise as probes of the structure of dense hadronic matter formed in
relativistic heavy ion collisions?  Although the final answer to this 
question is still extant, several conclusions can be drawn now.  Clearly, 
the original expectation that any significant reduction of $J/\psi$ 
formation in nuclear collisions would signal the creation of a quark-gluon 
plasma had to be revised.  We now understand that the suppression observed 
with light nuclear projectiles at the CERN-SPS is a smooth extrapolation of
the less-than-linear target mass dependence found in $p+A$ reactions.
We also know a credible mechanism for this suppression effect:  the
color-octet formation model.  However, it needs to be stressed that we
have almost no direct evidence for the color-octet $(c\bar c)$ state,
and we lack a quantitative description of the transition of the
color-octet pair to a color-singlet charmonium state from first principles.  
For these reasons, the color-octet model must be considered a reasonable, 
even likely, explanation of the suppression from $p+A$ to S + U reactions,
but it is far from being firmly and quantitatively established.

The available experimental evidence, combined with theoretical
arguments, points toward the emergence of novel $J/\psi$ suppression
mechanism in Pb + Pb collisions.  Because no experiments with projectiles 
between $^{32}$S and $^{208}$Pb have been done, it is unclear whether this 
new mechanism sets in suddenly or gradually.  However, the magnitude of 
the effect is such that it can hardly be explained as the absorption of 
color-singlet $J/\psi$ mesons on hadronic comovers, unless one wants to 
invoke the existence of hadrons, as we know them, at densities exceeding 1
fm$^{-3}$.  Moreover, a sizable comover suppression effect should be
visible already in S + U collisions, if the ``anomalous'' suppression
seen in the Pb + Pb system could be accounted for by absorption on
hadrons.  On the basis of the existing experimental evidence, the
structure of comoving matter has to undergo a significnt change between
S + U and Pb + Pb, if comover absorption is at the origin of the
suppression effect in Pb + Pb collisions.

In the following, we enumerate some weaknesses of the present arguments
regarding the ``anomalous'' $J/\psi$ suppression as evidence for the
creation of a quark-gluon plasma in Pb + Pb collisions at CERN.  We also
list some issues warranting further experimental and theoretical study.

\begin{enumerate}

\item The argument that $J/\psi$ suppression in S + U collisions is
``normal'' but that seen in Pb + Pb collisions has a novel, ``abnormal''
component relies critically on the precision of the extrapolation of the
$p+A$ results.  As the advocates of hadronic comover models have pointed
out, the Pb + Pb data would lose their cogency if the new suppression
mechanism were found to set in gradually as the nuclear projectiles get
heavier.  How large is the error in the value $\sigma_{c\bar c N}^{\rm
(abs)} = 7.3$ mb, i.e. how uncertain is the extrapolation of the
Gerschel-H\"ufner line to nuclear systems?  The extrapolation from $p+A$
to nuclear collisions depends critically on the assumption that the time
for color neutralization is sufficiently long, so that the $(c\bar c)$
pair remains in the color octet state while the nuclear matter sweeps
by.  Can one really exclude $\sigma_{c \bar c N}^{\rm (abs)} = 5$ mb if 
all errors in the $p+A$ data and the extrapolation to nuclear systems are
taken into account?

\item The viability of the hadronic comover model critically depends on
the value of the absorption cross section for $J/\psi$ on hadrons.  It
is possible to obtain a safe upper bound for $\sigma_{\psi h}^{\rm (abs)}$ 
in a hadronic gas of imprecisely known composition?  A systematic study of
$J/\psi$ absorption within effective meson theories including mesons
containing heavy quarks would be extremely useful.  Another important
study would include heavier hadronic states in the operator-product
expansion apaproach to $J/\psi$ absorption.  Since $\langle h\vert
\alpha_s E^2\vert h\rangle \sim m_h^2$, heavier mesons can act as ample
sources of gluons that may excite and dissociate the $J/\psi$.  (The
higher dissociation efficiency of $\rho$-mesons is also visible in the
D-meson exchange model, see section III.C.)  For example, it would be
interesting to explore the $J/\psi$ dissociation rate in a resonance gas
with a Hagedorn-type excitation spectrum.

\item An important issue that has not been investigated extensively
is the problem of $\chi_c$ absorption \cite{Heinz}.  Clearly, 
$\sigma_{\chi h}^{\rm abs}$ must lie somewhere between the cross sections 
for $J/\psi$ and $\psi'$ absorption. The question is if and where 
``anomalous'' $\chi_c$ absorption sets in. Can hadronic comover absorption
of $\chi_c$ in S + U be ruled out? Can the ``anomalous'' absorption of
$J/\psi$ in Pb + Pb be attributed entirely to simiarly ``anomalous''
absorption of the component due to feed-down from $\chi_c$?  Vogt's 
extensive study \cite{Vogt97} seems to indicate that it is difficult to 
find a scenario including feed-down from $\chi_c$ to $J/\psi$ that is 
compatible with all existing heavy ion data.

\item There are many improvements that could be made to microscopic
models of $J/\psi$ suppression.  Including a realistic energy dependence
of $\sigma_{\psi h}^{\rm (abs)}$, similar to the one found in section
III.C, would probably lead to an effective nonlinear dependence of the
comover effect on $dN/dy$, because higher comover densities are usually
correlated with increased average kinetic energy per particle.

Because the $p_T$-spectrum of $J/\psi$-suppression will be used
as a tool for discriminating between different mechanisms, a study of
final state changes in the $p_T$-distribution of $J/\psi$, e.g. due to
elastic scattering on comovers, would be interesting.  In more general
terms, microscopic transport models could be used to systematically
assess kinematical effects that can invalidate the Glauber approximation
in the context of $J/\psi$ production, in particular, effects of
transverse expansion and changes in the $p_T$ distribution of the
$J/\psi$ by elastic scattering in the medium \cite{Rafelski}.  Both
effects increase for heavier nuclei.

\item Desirable improvements in the experiment include a reduction of
statistical and systematic errors by taking more data.  Higher
statistics data would also permit the determination of the $J/\psi$
suppression factor in a variety of $E_T$ bins and over a range of values
of $p_T$, the transverse momentum of $J/\psi$.  A systematic study of
the mass dependence of the effect for several symmetric systems from S +
S to Pb + Pb would be especially important, so that the specific
predictions of the deconfinement model can be checked against the data.
An excitation function would also be of interest, but due to the steep
beam-energy dependence of $J/\psi$ production it may be difficult to
eliminate systematic errors with the required accuracy.  As pointed out
by Kharzeev and Satz \cite{KS95a}, the inverse kinematic experiment 
(Pb + S) would allow for a test of formation time effects on the 
suppression factor.

\item The comparison of the experimental data with theoretical models
will be facilitated, if the data are plotted against measured variables 
(such as $E_T$) rather than model dependent derived quantities (such as
$L$).  Of course, a detailed understanding of the experimental trigger 
conditions is indispensible for a meaningful comparison between
experiment and theory.

A clear set of ground rules needs to be established for the data
analysis as well as for the comparison with theory.  If the ratio between
$J/\psi$ yield and Drell-Yan (DY) background is plotted, only those mass
ranges should be included where the DY process can be unambiguously
identified.  It makes little sense to include the region under the
$J/\psi$ itself, where the DY pairs make only a small contribution.

Theorists also need to take a careful look at the soundness of
their models.  While microscopic transport models may have the
appearance of great realism, their quantitative results can depend on
many implicit model assumptions that are difficult to analyze.  Thus,
more schematic models will continue to provide useful analytic tools.
However, blind or ad-hoc parameter fitting can easily confuse the
situation.  Parameter values must be related to other experimental data,
providing support for the fit, and should not exceed the limits of
applicability of the schematic model itself.  In some cases, it may be
possible to check by {\it ab initio} QCD calculations whether a certain
parameter set makes sense.  As the picture of $J/\psi$ suppression in
nuclear reactions takes on a firmer shape, models must increasingly be
judged by their ability to predict new observables quantitatively.

\end{enumerate}

Finally, with the experimental emphasis shifting from the CERN-SPS to
RHIC in the years ahead, it may be possible to observe not only the
nuclear suppression of the $J/\psi$, but also the suppression of the
$\Upsilon$ in heavy collision systems.  Parton cascade models predict
initial energy densities of the order of 50 GeV/fm$^3$, far exceeding
the deconfinement threshold of the $(b\bar b)$ ground state.  The
advantage of the $\Upsilon$ would be that its suppression in $p+A$
collisions is much weaker than that of the $J/\psi$ and, therefore, its
almost complete suppression in a deconfined phase would be quite
spectacular.

\subsection*{Acknowledgements:}

This review would not have been written without the encouragement of 
Carlos Louren\c co, who provided the challenge with the invitation to
present a lecture in the CERN Heavy Ion Forum.  
I also thank Sergei Matinyan for many illuminating discussions.  
This work was supported in part by a grant from the U.S. 
Department of Energy DE-FG02-96ER40945.

\section*{Appendix A: $J/\psi$ Dissociation in a Medium}

Here we consider the dissociation of a $J/\psi$ state in a medium within
the Bhanot-Peskin formalism.  The fundamental idea is that the
dissociation is initiated by the absorption of a gluon from the medium
which excites the color-singlet $(c\bar c)$ pair into a color-octet
continuum state $\vert (c\bar c)^{(8)}, \epsilon\rangle$, where
$\epsilon$ denotes the final state energy.  The S-matrix element is
\begin{equation}
S_{fi} = {1\over i\hbar} \int_{-\infty}^{\infty} dt \langle (c\bar
c)^{(8)} \epsilon \vert g\vec r\cdot \vec E(r,t)\vert (c\bar c)^{(1)}
\rangle. \label{A1}
\end{equation}
We will make use of the dipole approximation assuming that the
dissociation is dominated by long-range gluonic interactions.  In the
heavy quark limit, the binding energy of the $(c\bar c)^{(1)}$ state is
${4\over 9}\alpha_s^2m_Q$, whereas its inverse size is $a^{-1} =
{2\over 3}\alpha_s m_Q$.  When $\alpha_s \ll 1$, the characteristic
wavelength of a gluon that is energetically capable of dissociating the
$J/\psi$ state is larger than the $J/\psi$ radius.  This approximation
is somewhat marginal for the $J/\psi$ state; it should be very good for 
the $\Upsilon$.  After performing the SU(3)-color algebra, the dipole matrix
element is given by
\begin{equation}
{g\over \sqrt{6}} \int d^3r \varphi_8^{\epsilon}(r) \vec r \varphi_1(r),
\label{A2}
\end{equation}
where $\varphi_1(r)$ and $\varphi_8^{\epsilon}(r)$ are the spatial
wavefunctions of the singlet and octet states, respectively.  We neglect
recoil effects, and we use a Coulombic $1s$-wavefunction for the singlet
state and a plane wave for the octet state.  The resulting S-matrix then
takes the form
\begin{equation}
S_{fi} = {g\over \sqrt{6}} \sqrt{{\pi a^3\over V}} 32 a^2 {\vec
E^a(\omega)\cdot\vec p \over (1+p^2a^2)^3}. \label{A3}
\end{equation}
Here $\vec E^a(\omega)$ is the spectral distribution of gluon fields in
the medium at frequency $\omega= \epsilon_8-\epsilon_1$, $a$ is the Bohr 
radius of the $(c\bar c)^{(1)}$ state, $\vec p$ is the relative momentum 
of the $(c\bar c)^{(8)}$ pair, and $V$ denotes the quantization volume.
Integrating over $\vec p$ and assuming color neutrality of the medium,
i.e.
\begin{equation}
\langle E_i^a(\omega) E_j^b(\omega)\rangle = {1\over 24} \delta_{ij}
\delta_{ab} \langle \vert E (\omega)\vert^2\rangle, \label{A4}
\end{equation}
where $i,j$ denote spatial vector and $a,b$ color indices, we finally
obtain the transition probability
\begin{equation}
P_{if} = {2\over 3} \pi\alpha_sa^2 \; \langle \vert E
(\omega)\vert^2\rangle. \label{A5}
\end{equation}
Since $\langle \vert E(\omega)\vert^2\rangle$ is proportional to the
total interaction time $T_{fi}$, the result (\ref{A5}) corresponds to a
constant dissociation rate.

The color-electric power density ${1\over T_{fi}} \langle \vert
E(\omega)\vert^2\rangle$ of the medium can be evaluated analytically
for a variety of simple models.  One such model is a dilute gas of color
charges with a given density and momentum distribution.  Denoting the
Casimir of the color charge by $Q^2$, such a model yields
\begin{equation}
{1\over T_{fi}} \langle \vert E(\omega)\vert^2\rangle \approx {\pi\over
2}\alpha_s Q^2 \tilde\rho(\omega), \label{A6}
\end{equation}
where $\tilde\rho (\omega)$ is a $\omega$-dependent, weighted average of
the density of charges in the medium.

If the medium is represented as a dilute gas of hadrons, the power
spectrum is related to the gluon structure functions of the hadrons
$G_h(x)$ where $x=\omega/p_h$, in the spirit of the calculation of
Kharzeev and Satz \cite{KS94}.  Given a momentum spectrum of hadrons
$f(p_h)$ and chemical abundances of different hadrons, the power
spectrum can be evaluated.

Crude estimates of the dissociation rate in these two models yield
rather long survival times of the $J/\psi$ in any realistic hadronic
medium.  For example, counting the valence quarks in mesons and
combining (\ref{A5}) with (\ref{A6}), would yield a dissociation rate
\begin{equation}
\Gamma_{\rm dis} \approx {8\over 9}\pi^3 \alpha_s^2 a^2 \tilde\rho_{\rm
meson}, \label{A7}
\end{equation}
where only sufficiently energetic mesons are counted.

\section*{Appendix B:  Matrix Elements for Charm Exchange}

In this Appendix we list the explicit expressions for the invariant
matrix elements for the charm exchange reactions described by the
Feynman diagrams in Fig.~\ref{fig9}.  We begin with the two diagrams for
$J/\psi$ absorption on pions (diagrams \ref{fig9}a,b).  Averaging over initial
and summing over final isospin and spin states, we obtain
\begin{equation}
\overline{\vert M_a\vert^2} = {8\over 3}g_{\psi DD}^2 g_{\pi DD^*}^2 
{\left(m_{\pi}^2 - {(p_{\pi}\cdot p_{D^*})^2\over m_{D^*}^2}\right) 
\left(m_D^2-{(p_{\psi}\cdot p_D)^2\over m_{\psi}^2}\right) 
\over (q^2 - m_D^2)^2}
\label{B1}
\end{equation}
where $q=p_{\pi}-p_{D^*} = p_D-p_{\psi}$ is the momentum transfer of the
reaction.  Introducing the Mandelstam variables $s,\;t$ and $u$, and the
shorthand notation $t'=t-m_D^2$, we obtain
\begin{equation}
\overline{\vert M_a\vert^2} = {1\over 6} g_{\psi DD}^2 g_{\pi DD^*}^2
{1\over t^{'2}} \left( 4m_{\pi}^2 - m_{D^*}^2 + 2t' - {t^{'2}\over
m_{D^*}^2}\right) \left( 4m_D^2 - m_{\psi}^2 + 2t' - {t^{'2}\over
m_{\psi}^2}\right). \label{B2}
\end{equation}
The result for the diagram (\ref{fig9}b) is identical, i.e.
\begin{equation}
\overline{\vert M_b\vert^2} = \overline{\vert M_a\vert^2}. \label{B3}
\end{equation}
The contributions of the diagrams (\ref{fig9}c,d) need to be added coherently,
because they lead to identical final states.  However, their results are
related by crossing symmetry $(t\leftrightarrow u)$.  After somewhat
lengthy algebra one finds:
\begin{eqnarray}
\overline{\vert M_c+M_d\vert^2} &= &\frac{1}{18} g_{\psi DD}^2 
g_{\rho DD}^2 \left[ \frac{1}{t^{'2}} \left( 4m_D^2 -
\frac{(m_{\rho}^2-t')^2}{m_{\rho}^2}\right) \left(4m_D^2 -
\frac{(m_{\psi}^2-t')^2}{m_{\psi}^2}\right) \right.  \nonumber \\
&&+ \frac{1}{u't'} \left( 2s-4m_D^2 -
\frac {(m_{\rho}^2-t')(m_{\rho}^2-u')}{m_{\rho}^2} \right)
\left(2s - 4m_D^2 - \frac{(m_{\psi}^2 - t')(m_{\psi}^2-u')}{m_{\psi}^2}
\phantom{1\over t^2} \right) \nonumber \\
&&+ \left. (t' \leftrightarrow u') \right] \label{B4}
\end{eqnarray}
where again $t'=t-m_D^2$, $u'=u-m_D^2$.  Of course, the crossing related
diagrams (\ref{fig9}c,d) yield the same contribution to the total absorption
cross section, with an additional interference term.

Finally, for completeness, we give the expression for the decay width of
$D^*\to D\pi$:
\begin{equation}
\Gamma_{D^*} = {g_{\pi DD^*}^2\over 8\pi} {p_{\pi}^3\over m_{D^*}^2}.
\label{B5}
\end{equation}

\end{document}